\documentclass[aps,prl,twocolumn,superscriptaddress,longbibliography]{revtex4-2}

\usepackage{mathrsfs}
\usepackage{tabularx}
\usepackage{slashed}
\usepackage{amsmath}
\usepackage{amsfonts}
\usepackage{physics}
\usepackage{bm}
\usepackage{graphicx,color}
\usepackage{times}
\usepackage[caption=false]{subfig}
\usepackage[colorlinks,linkcolor=red,citecolor=blue]{hyperref}

\begin{document}

\title{Quantum Anomalous Hall Crystal at Fractional Filling of Moir\'e Superlattices}
\author{D. N. Sheng}
\email{donna.sheng1@csun.edu}
\affiliation{Department of Physics and Astronomy, California State University Northridge, Northridge, California 91330, USA}
\author{Aidan P. Reddy}
\email{areddy@mit.edu}
\affiliation{Department of Physics, Massachusetts Institute of Technology, Cambridge, Massachusetts 02139, USA}
\author{Ahmed Abouelkomsan}
\email{ahmed95@mit.edu}
\affiliation{Department of Physics, Massachusetts Institute of Technology, Cambridge, Massachusetts 02139, USA}
\author{Emil J. Bergholtz}
\email{emil.bergholtz@fysik.su.se}
\affiliation{Department of Physics, Stockholm University, AlbaNova University Center, 106 91 Stockholm, Sweden}
\author{Liang Fu}
\email{liangfu@mit.edu}
\affiliation{Department of Physics, Massachusetts Institute of Technology, Cambridge, Massachusetts 02139, USA}
\date{\today}
\begin{abstract} We predict the emergence of a state of matter with intertwined ferromagnetism, charge order and topology in fractionally filled moir\'e superlattice bands. Remarkably, these {\it quantum anomalous Hall crystals} exhibit a quantized integer Hall conductance that is different than expected from the filling and Chern number of the band. Microscopic calculations show that this phase is robustly favored at half-filling ($\nu=1/2$) at larger twist angles of the twisted semiconductor bilayer $t$MoTe$_2$.  
\end{abstract}

\maketitle

Moir\'e superlattices provide an exciting venue for realizing correlated and topological quantum states due to strong interaction effect in minibands \cite{andrei_graphene_2020,mak_semiconductor_2022}. As a common manifestation of electron correlation, a variety of symmetry breaking phenomena have been observed in graphene and semiconductor moir\'e materials. These include superconductivity \cite{caoUnconventionalSuperconductivityMagicangle2018}, spin/orbital ferromagnetism \cite{sharpe2019emergent, anderson2023programming} and Wigner crystallization \cite{li2021imaging, jinStripePhasesWSe22021}, which spontaneously break gauge symmetry, flavor symmetry and lattice translation symmetry respectively. Another exciting aspect of moir\'e physics lies in topological quantum phenomena, which are characterized by quantized transport and thermodynamics responses. 
These include the quantum Hall effect arising from graphene Hofstadter band at strong magnetic field \cite{dean_hofstadters_2013, spanton2018observation}, and the recently observed quantum spin Hall effect in semiconductor moir\'e band at zero field \cite{kang_observation_2024,kang2024observation2}.   

Moir\'e quantum matter may also exhibit intertwined symmetry breaking and topological phenomena. One example is the quantum anomalous Hall (QAH) states observed in graphene and semiconductor moir\'e superlattices exhibit integer quantized anomalous Hall effect at zero magnetic field \cite{serlin2020intrinsic,li2021quantum, foutty2023mapping}. 
Even more remarkably, the fractional quantum anomalous Hall (FQAH) states have recently been observed 
for the first time in twisted semiconductor bilayer $t$MoTe$_2$ and pentalayer rhombohedral graphene/boron nitride moir\'e heterostructure \cite{park2023observation,  xu_observation_2023, cai2023signatures, zeng2023thermodynamic, lu_fractional_2024}.  

In twisted semiconductors, the existence of QAH and FQAH states was theoretically predicted \cite{devakul2021magic, li2021spontaneous, crepel2023anomalous,abouelkomsan2020particle,repellin2020chern} as a consequence of spin ferromagnetism spontaneously breaking time reversal symmetry as well as topological moir\'e band with spin-contrasting Chern numbers \cite{wu2019topological}. At fractional band filling, ferromagnetism lifts the spin degeneracy, and Coulomb interaction drives spontaneously spin polarized electrons in Chern band into a fractional Chern insulator state---the lattice analog of the fractional quantum Hall state \cite{neupert2011fractional,sheng2011fractional,regnault2011fractional, LIU2024515}.  Importantly, ferromagnetism and topological order both arise from the same electrons in moir\'e band, and therefore are strongly intertwined.

In this work, we present a novel type of topological states intertwined with both time-reversal and translation symmetry breaking in moir\'e superlattices. Specifically, we uncover the emergence of integer QAH effect in a commensurate charge density wave (CDW) state at fractional filling of the moir\'e unit cell. 
Focusing on twisted semiconductor bilayer $t$MoTe$_2$, by exact diagonalization (ED) study and many-body Chern number calculation, we find   
a novel QAH state with $2\times 2$ CDW at half filling ($\nu=1/2$) in a previously unexplored twist angle  
range. This state features a quantized Hall conductance $\sigma_{xy}=1$ in units of $e^2/h$, which differs from the underlying band Chern number multiplied by the filling factor (as in FQH states in a Landau level):   
$\sigma_{xy}\neq \nu C_{\rm band}$.   
The existence of QAH effect in a CDW state defines a novel quantum phase of matter that intertwines ferromagnetism, charge order and topology. Such state was recently termed a QAH crystal \cite{song2023}.

Historically, the possible coexistence of quantized Hall effect and charge density wave was theoretically considered for two-dimensional electrons in a strong magnetic field \cite{halperin1986, kivelson1986,kivelson1987,halperin1989,ganpathy2000}. Recent advances on moir\'e materials have stimulated interest in QAH crystals. While the original ``Hall crystal'' breaks continuous translation symmetry and therefore exhibits a gapless neutral mode \cite{halperin1989,ganpathy2000}, QAH crystals in a moir\'e lattice setting break discrete symmetry and therefore are not expected to have gapless excitations. Evidence for ``symmetry-broken Chern insulators'' has been observed 
in several systems, including  magic-angle graphene aligned with boron nitride and twisted monolayer-bilayer graphene. While these states are consistent with broken lattice translation symmetry, their quantized Hall effect was found at finite magnetic field.
Theoretical studies have also explored the possibility of (zero-field) QAH crystals  in twisted multilayer graphene at half-integer fillings \cite{wilhelm2021interplay, Wilhelm_2023} as well as  
semiconductor moir\'e materials at $\nu=2/3$ \cite{pan_topological_2022, xu2024maximally, song2023}. Despite significant progress, the QAH crystal has not yet been experimentally found in any system, and to our knowledge, the possibility of QAH crystals at $\nu=1/2$ 
 with $\sigma_{xy}\neq \nu C_{\rm band}$ has not been considered before.  

Twisted bilayer semiconductors host time-reversed pairs of topological moir\'e bands, which feature opposite Chern numbers in opposite spin/valley sectors: $C_\uparrow= - C_\downarrow =1$ \cite{wu2019topological}.  
At small twist angle, the lowest hole band is sufficiently narrow, which favors strong interaction effect \cite{devakul2021magic}. 
Previous theoretical studies on fractional-filling states have focused on $\theta<4^\circ$ where FQAH states have been experimentally observed \cite{reddy2023fractional, wang2023fractional, goldman2023zero, dong2023composite, xu2024maximally, Yu2023, abouelkomsan2024band, reddy2023toward}.   

As a function of twist angle, the dispersion and wave function of low-energy moir\'e bands change considerably. At  small twist angles, the density of states is mostly concentrated at MX and XM stacking sites of each moiré unit cell, which collectively form a honeycomb lattice, thus leading to an effective Kane-Mele tight binding model description \cite{wu2019topological}. 
At larger twist angles, the local density of states is more uniform, but peaked at the triangular lattice of MM stacking sites where interlayer tunneling is strongest \cite{reddy2023fractional, qiu2023interaction}.  
Moreover, the bandwidth changes non-monotonously on the twist angle: it reaches a minimum at a magic angle and increases away from it \cite{devakul2021magic, morales2023magic}. 
The twist-angle-dependent character of the lowest band's wave function and dispersion play an important role in the competition between crystalline and liquid states at fractional fillings such as $\nu=1/3$ and $2/3$ \cite{reddy2023fractional, abouelkomsan2024band}. 
In the following we shall explore their consequence at $\nu=1/2$.

\begin{figure}
   \includegraphics[width=\linewidth]{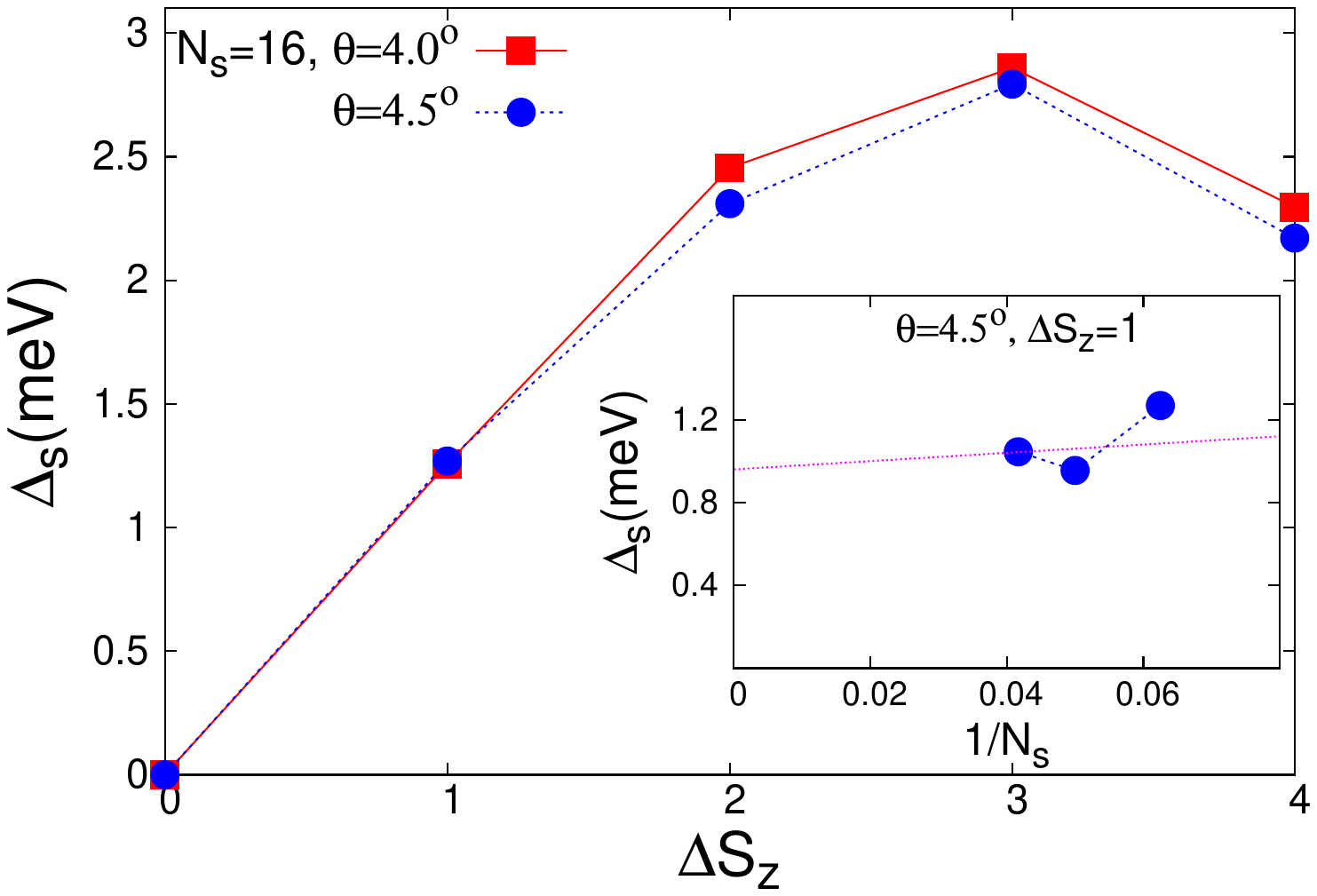}
\vspace{-1.3cm}
\caption{The spin gap $\Delta_s$ for $N_s=16$ system as a function of
the spin flip $\Delta S_z=S_{zmax}-S_z$ for twist angles $\theta=4.0^\circ$ and $4.5^\circ$.
In the inset, we show $\Delta_s$ for $N_s=16, 20$ and $24$ with $\theta=4.5^\circ$  at $\Delta S_z=1$.
} \label{fig:spin_gap}
\end{figure}

{\it Ferromagnetism} Using ED method,  we first study $t$MoTe$_2$ at the filling of $\nu=1/2$ holes per moir\'e unit cell. We consider a finite system with lattice vectors ${\bf L}_1, {\bf L}_2$,
with the number of moiré unit cells $N_1, N_2$ along these two  directions, respectively.
We mainly focus on twist angle $\theta \gtrsim 4^\circ$ where the lowest hole band is more dispersive and its Berry curvature is less uniform \cite{reddy2023fractional}. These features motivate us to explore the possibility of new quantum states beyond those of the lowest Landau level.  Since the numbers of spin-$\uparrow$ and $\downarrow$ holes---denoted as $N_\uparrow$ and $N_\downarrow$ respectively---are separately conserved, we perform ED calculation for all spin configurations $(N_\uparrow, N_\downarrow)$ with hole number $N_h=N_\uparrow + N_\downarrow=\nu N_s$, where $N_s=N_1\times N_2$ is the number of unit cells in the system. $S_z=(N_\uparrow-N_\downarrow)/2$ denotes the total spin $S_z$. 
Our ED calculation is carried out within the Hilbert space of the time-reversed pair of lowest bands. 

Using the continuum model Hamiltonian in Ref.\cite{reddy2023fractional} with parameters for $t$MoTe$_2$,  we compare energies of lowest states in different spin sectors and  determine the ground state spin polarization.  The spin gap $\Delta_s=
 E_{min}(S_z)-E_{min}(S_{zmax})$ as a function of 
spin-flip $\Delta S_z=S_{zmax}-S_z$  is shown in Fig.~\ref{fig:spin_gap} for the twist angle range of our interest, $\theta=4^\circ-4.5^\circ$, where a clear gap is shown  for all
spin sectors for $N_s=16$.  Furthermore,  as show in the inset of Fig.~\ref{fig:spin_gap}, 
spin gaps of $N_s=16$, 20 and 24 site clusters are robust,
and a finite-size scaling with system size $N_s$ shows that the spin gap remains finite for large system limit. 
Thus our results demonstrate  that under Coulomb interaction with dielectric constant $\epsilon=10$, the ground state at $\nu=1/2$ is fully spin polarized over a wide range of twist angles, up to at least $\theta=4.5^\circ$. Since the underlying band structure is strongly spin dependent due to spin-valley locking, the system lacks $SU(2)$ spin-rotational symmetry, thus leading to a finite spin gap in the ferromagnetic ground state \cite{crepel2023anomalous, reddy2023fractional, Yu2023}. In the rest of this work, we perform ED calculations within the Hilbert space of spin-$\uparrow$ holes in the lowest Chern band, which allows us to access much larger system sizes up to 32 sites.   

\begin{figure}
   \includegraphics[width=\linewidth]{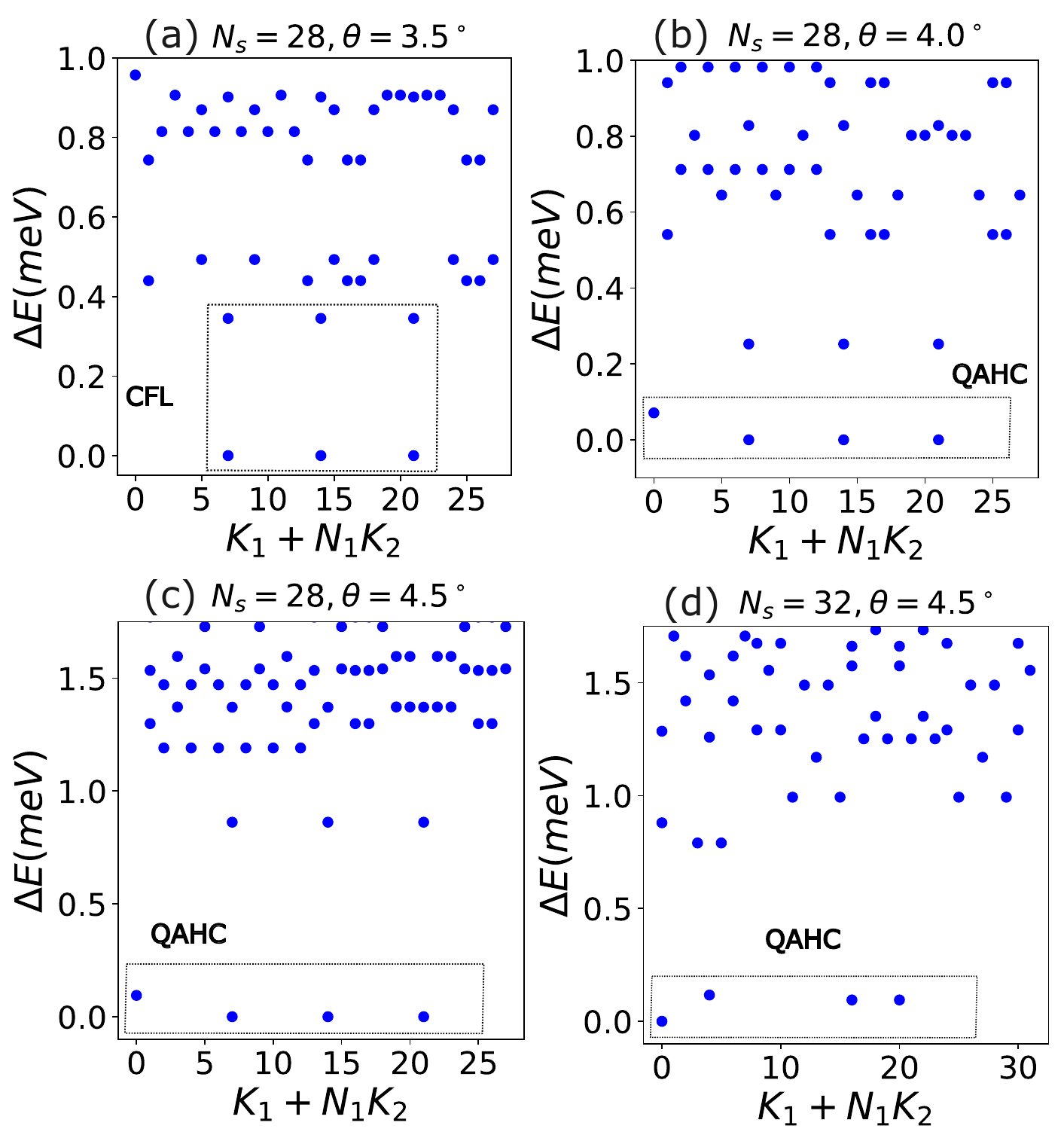}
\caption{Energy spectra evolution from
CFL to candidate Hall crystal states for (a) $\theta=3.5^\circ$, (b) $\theta=4.0^\circ$, (c) $\theta=4.5^\circ$ with $N_s=28$ and (d) $\theta=4.5^\circ$, $N_s=32$ at filling number
$\nu=1/2$. 
} 
\label{fig:energy_spectrum}
\end{figure}

{\it Charge density wave}  We explore the evolution of the ground state and low-lying states of $t$MoTe$_2$ at $\nu=1/2$ with the twist angle. 
Consistent with previous studies \cite{goldman2023zero,dong2023composite}, the ground state is a composite Fermi liquid (CFL) for a wide range of twist angles. Fig.~\ref{fig:energy_spectrum} shows the many-body energy spectra as a function of the momentum ${\bf k}=K_1{\bf T}_1+K_2{\bf T}_2$ with ${\bf T}_1, {\bf T}_2$  as unit vectors of crystal momentum for $\theta=3.5^{\circ}$, $4.0^{\circ}$ and $4.5^{\circ}$ with  $N_s=28$  cluster and $\epsilon=10$.  
We find a pair of quasidegenerate ground states with center-of-mass momenta at $M$, similar to the energy spectrum of the CFL state in half-filled lowest Landau level at $\theta=3.5^{\circ}$. 
As the twist angle further increases, however, a level crossing between CFL ground states and higher-energy states occurs around $\theta=4.0^{\circ}$, indicating a quantum phase transition into a new phase at larger twist angle.  

At $\theta=4.5^\circ$, on both $N_s=28$ and $32$ clusters, we find four nearly degenerate ground states with center-of-mass momentum at $\Gamma$ and three $M$ points, which are separated by an energy gap from higher-energy states (Fig.~\ref{fig:energy_spectrum}(c-d)). 
To understand the nature of the ground state, we study its projected density structure factors
\begin{align}
\begin{split}
    S(\bm{q}) = \frac{\langle {\bar{\rho}}(-\bm{q}){\bar{\rho}}(\bm{q})\rangle}{N_s} -
    \frac {\langle {\bar{\rho}}(0)\rangle^2} {N_s} \delta_{\bm{q},\bm{0}}
    \end{split}
\end{align}
and the pair correlation functions
\begin{align}
      g(\bm{r},\bm{r}')=\frac{\langle n(\bm{r})n(\bm{r}')\rangle -\delta(\bm{r}-\bm{r}')\langle n(\bm{r})\rangle}{(N_h/A)^2}
\end{align}
Here ${\bar{\rho}}_{\bm{q}} = P \sum_{i}e^{-i\bm{q}\cdot\bm{r}_i}P$ is the projected density fluctuation operator where $P$ is a projector onto the first moiré hole miniband.
The local density operator is $n(\bm{r}) = \sum_{l}\psi^{\dag}_{l}(\bm{r})\psi_{l}(\bm{r})$ where $\psi^{\dag}_{l}(\bm{r})$ creates a hole at position $\bm{r}$ in layer $l$.  $A$ is the area of the system.
We also take an average over Moir\'e unit cell 
 so that $g(\bm{r}-\bm{r'})$ only depends on $\bm{r}-\bm{r'}$.

\begin{figure}
   \includegraphics[width=1.0\linewidth]{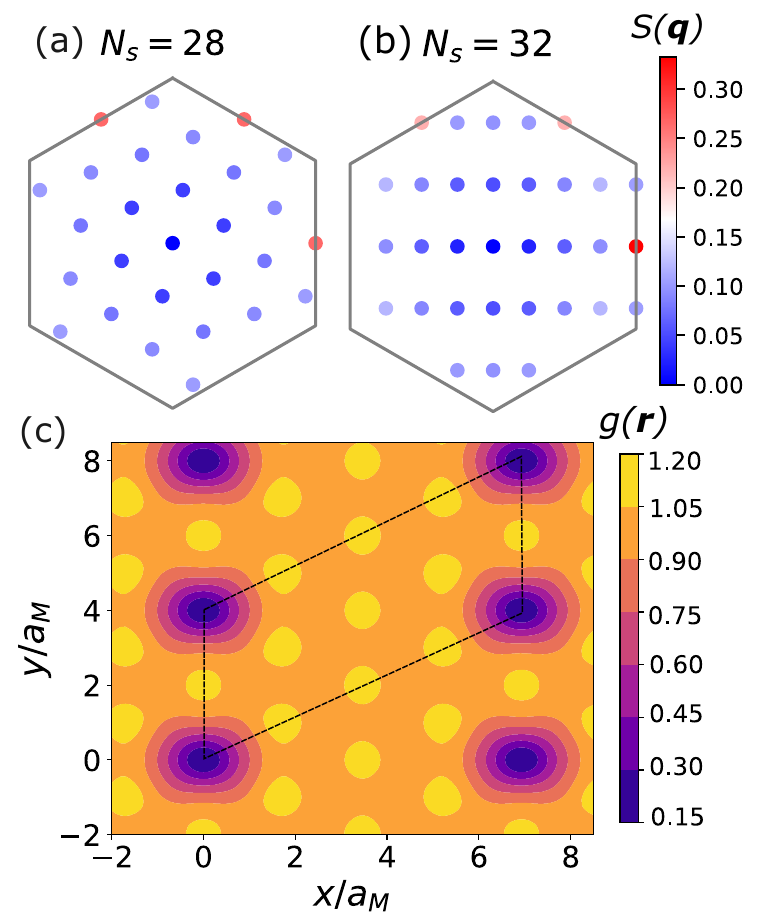}\\
\caption{(a), (b) Density structure factors for $N_s=28$ and $32$, respectively; (c) pair correlation function for $N_s=32$  at filling number $\nu=1/2$ at twist angle $\theta=4.5^\circ$. } 
\label{fig:Densq}
\end{figure}

As shown in Fig.~\ref{fig:Densq}, $S({\bm q})$ shows prominent peaks at three $M$ points and the peak intensity increases with the system size from $N_s=28$ to 32, which indicates the emergence of CDW order in the thermodynamic limit. 
To illustrate the real space order, we show pair correlation function $g(\bm r)$ (see Fig.~\ref{fig:Densq}), which demonstrates a $2\times 2$ CDW  consistent with the three $M$-point peaks in $S({\bm q})$. Strong peaks in $g(\bm r)$ are found on $1/4$ of MM sites on the moir\'e superlattice, where the local density of states of the moir\'e band is high. Importantly, the ordered $2\times2$ structure in the pair correlation extends throughout the entire system of $N_s=32$ sites---the largest size accessible in our  ED study, and similar results are found for
$N_s=24$ and $28$.  Notably, even though the 32 site cluster geometry is not threefold symmetric as the 28 site cluster is, the pair correlation function here still shows the $2\times 2$ CDW pattern. This strongly suggests that the CDW state at $\nu=1/2$ has a bidirectional $2\times 2$ ordered structure, rather than a unidirectional $2\times 1$ stripe. 
By performing similar calculations for various twist angles, we find that for $\epsilon=10$, the $2\times 2$ CDW state at $\nu=1/2$ exists for twist angles in the range $4.0^\circ\lesssim \theta \lesssim 5.0^\circ$.

{\it Many-body Chern number} To probe the topological order of this CDW state, we further calculate the many-body Chern number as an integral invariant of many-body wave function over twist boundary condition\cite{sheng2003chern,sheng2011fractional}. 
The Chern number is defined as,
\begin{equation}
    \mathcal{C} 
 ={i\over 2\pi}\int d\phi_1 d\phi_2
\{\langle { \partial \psi \over \partial \phi_1 } |{\partial \psi
 \over
\partial \phi_2}\rangle - c.c.
\},
\end{equation}
where $|\psi\rangle$  is the many-body state and the integral is over the $2\pi\times 2\pi$ boundary phase space with  twist boundary phases $\phi_1$ and $\phi_2$ along lattice vectors $\bm{L}_1$ and $\bm{L}_2$ directions, playing the role of magnetic fluxes.  The twist angles shift the
kinetic momentum of each particle  by $\frac {\phi_1} {2\pi}{\bm {T}_1}+ \frac {\phi_2 }{2\pi} {\bm {T}_2}$.  We discretize the boundary phase space into at least $12\times 12$ square meshes  and numerically obtain total Berry phase summing over  each square. We calculate the consecutive  wave function overlaps $\langle \psi(\phi_1,\phi_2)|\psi(\phi'_1,\phi'_2)\rangle$ 
around each square, which gives the Berry phase\cite{okamoto2022top} from both the single particle basis states 
of the moir\`e band and  the many-body contribution. 
Energy spectrum flow with inserted flux always shows finite gaps between different energy levels within the same momentum sector,  which  warrants  a well defined Chern number for many-body states  (see SM\cite{SM} for more details).  
 For such many-body systems, the boundary phase averaged
Hall conductance can be obtained as $\sigma_{xy}= C$\cite{sheng2003chern}.

\begin{figure}
   \includegraphics[width=1.0\linewidth]{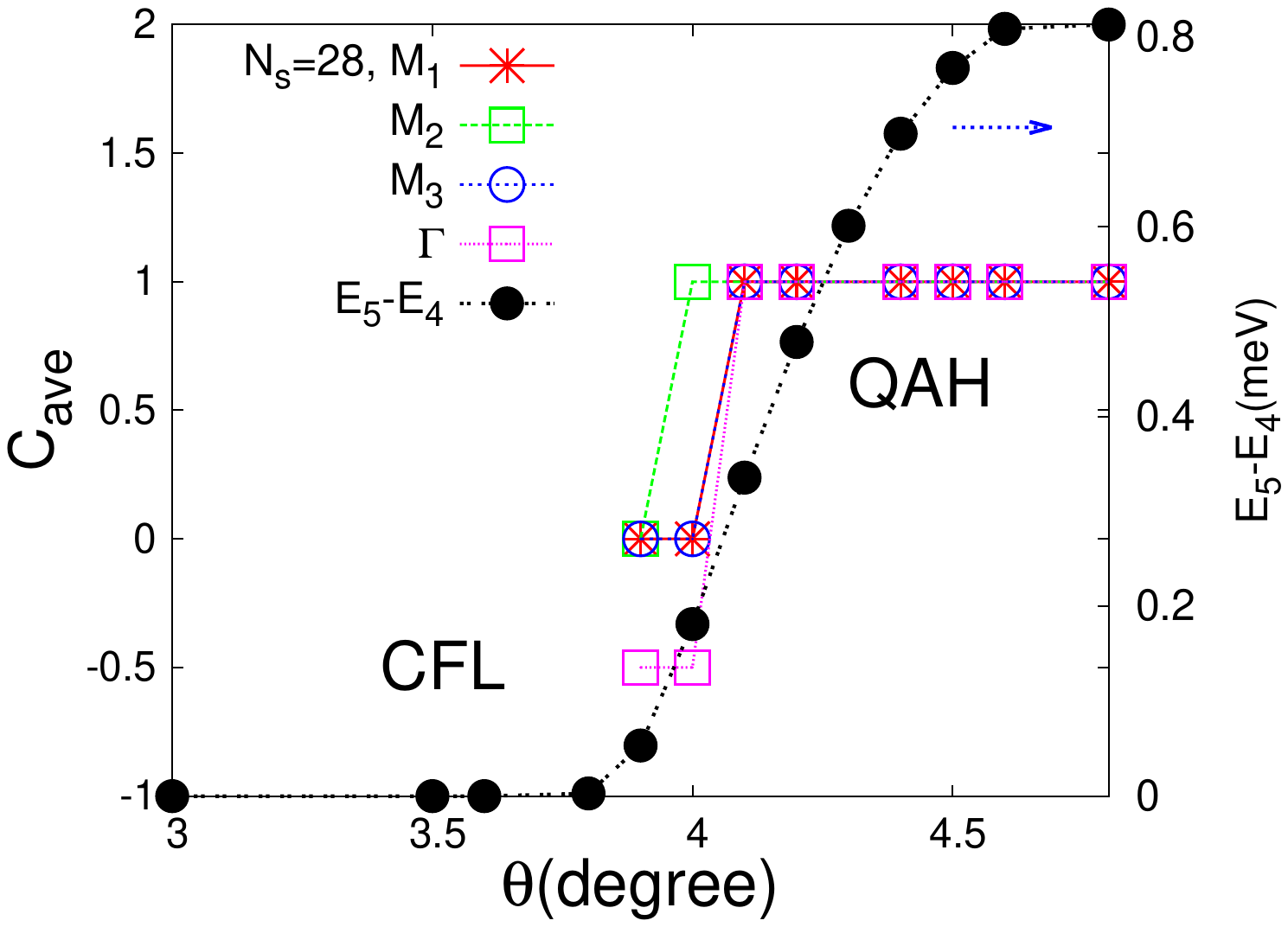}
\vspace*{-1.0cm}
\caption{Left axis: Chern numbers averaged over each member of nearly degenerate lowest energy states, as well as the first excited state in the same momentum sector, versus $\theta$ for $N_s=28$ system at filling $\nu=1/2$. 
Right axis: Excitation gap $E_5-E_4$ between the fifth and fourth lowest energy levels.
} 
\label{fig:chern_number}
\end{figure}

We focus on larger twist angle $\theta$ and identify  the Chern numbers 
for two lowest energy states  
to characterize the topological nature of  the crystal state. 
As shown in Fig.~\ref{fig:chern_number} for a system of $N_s=28$ sites,
we demonstrate an integer quantization 
$C_{ave}=1$  for a QAH phase with $\sigma_{xy}=1$ at larger
$\theta \gtrsim 4.0^\circ$ side\cite{SM}. 
To determine the quantum phase transition between the  CFL and the crystal phase, we show the energy
gap $E_5-E_4$, which separates the lowest four levels from the next excited level.
Clearly we find this gap is identically zero for CFL regime for $N_s=28$, which grows monotonically
for larger $\theta$ into  the crystal phase, demonstrates a robust 4-fold degenerate ground state
and a robust excitation gap for the latter phase.  The transition 
is accompanied by Chern number fluctuations around the twist angle $\theta_c\sim 3.9^{\circ}-4.0^{\circ}$.

Remarkably, this topological phase transition from a compressible CFL phase to an incompressible QAH phase coincides with the symmetry breaking transition, i.e., the emergence of $2\times 2$ CDW order as seen from the structure factor and pair correlation\cite{SM}. 
We have also found that, for the QAH crystal phase,  both the ground state and the 1st excited state have the same integer quantized $C=1$ demonstrating the robustness of the topological quantization. Near the phase boundary at $\theta=4.0^\circ$, the ground state already has $C=1$  (see additional results for $N_s=24-32$ in the SM\cite{SM}). Combining all these findings, we conclude that at larger twist angle, the ground state of $t$MoTe$_2$ at $\nu=1/2$ is a QAH crystal that exhibits spontaneous ferromagnetism, $2\times 2$ CDW order, and quantized anomalous Hall effect $\sigma_{xy}=1$. This state exemplifies a topological quantum phase possessing  multiple symmetry-breaking order parameters.    
We also present numerical evidence for a QAH crystal with $\sigma_{xy}=1$  at filling number $\nu=3/4$ in the SM\cite{SM}. 

The existence of a $2\times 2$ CDW at $\nu=1/2$ with a quantized anomalous Hall conductance $\sigma_{xy}=1$  is unexpected and seems counterintuitive. For comparison, we note that CDW states were previously found at $\nu=1/2$ in semiconductor moir\'e superlattices with topologically trivial bands \cite{li2021imaging,jinStripePhasesWSe22021}. However, those CDW states are generalized Wigner crystals with $2\times 1$ stripe order, which host one charge per enlarged unit cell. Such Wigner crystals arise from strong Coulomb interaction between charges in a triangular lattice and are insulating states with zero Hall conductivity. Our QAH crystal state in the half-filled topological band of $t$MoTe$_2$ is clearly different. 

\begin{figure}
   \includegraphics[width=1.0\linewidth]{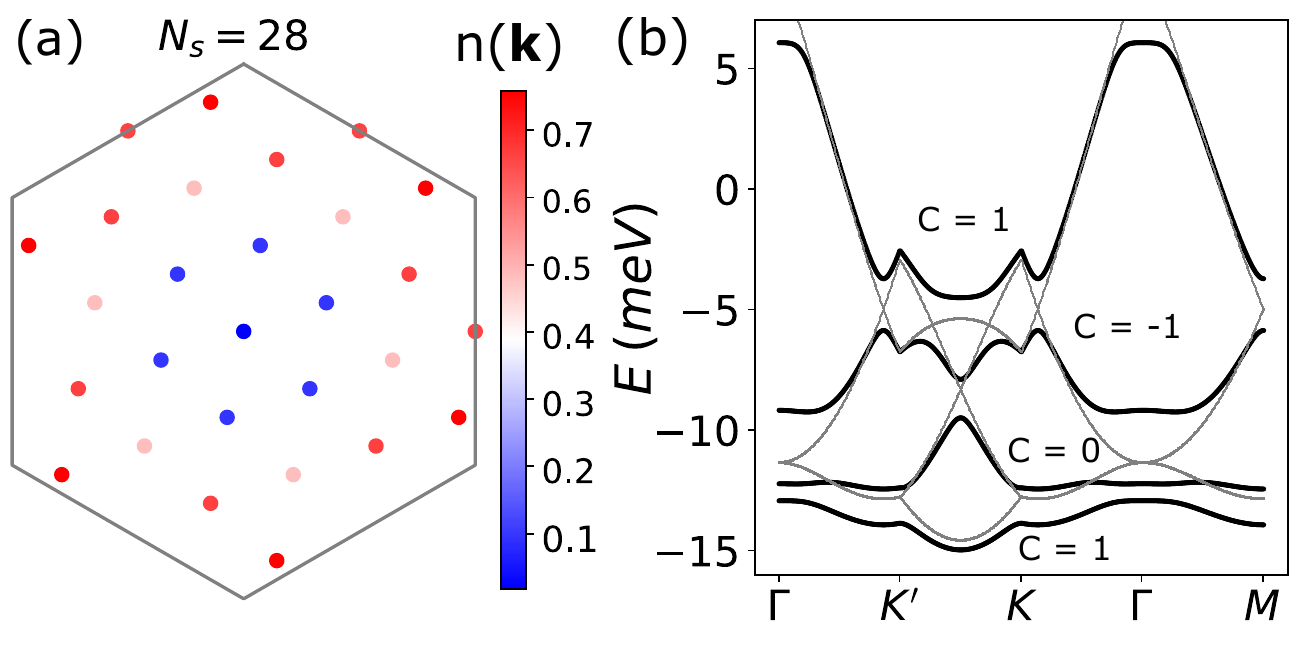}
\caption{The  momentum space occupation number obtained from ED
for $N_s=28$ at $\theta=4.5^\circ$. (b) The energy band splitting and Chern numbers in the
phenomenological  model with weak potential $V_0 = 1$~meV.    } 
\label{fig:pheno}
\end{figure}

To better understand the origin of the QAH crystal state, we calculate its momentum space occupation number $n({\mathbf{k}})$  at $\theta=4.5^\circ$,  shown in Fig.~\ref{fig:pheno}(a) for $N_s=28$ site cluster. Notably, $n({\mathbf{k}})$ is highly nonuniform: it is close to $0$ near $\gamma$ (hole band maxima)  and close to $1$ near  $\kappa, \kappa'$ (hole band mimima). This behavior contrasts with the nearly uniform momentum distribution previously found in the CFL state at smaller twist angles \cite{goldman2023zero}, but shares more similarity with $n({\mathbf{k}})$  of the noninteracting metal at $\nu=1/2$. This motivates us to introduce a phenomenological ``mean-field'' description for the $2\times 2$ CDW state. Our mean-field Hamiltonian consists of the noninteracting continuum model of t$\text{MoTe}_2$ that describes the kinetic energy of holes, as well as  an effective potential with $2\times2$ periodicity that represents the CDW order parameter: 
$V_{2\times2}(\mathbf{r}) = -2V_0  \sum_{i=1}^3 \cos(\mathbf{M}_i \cdot \mathbf {r})$
where $\mathbf{M}_i$ are the three $M$ points. Because of the enlarged real-space periodicity, the original hole band folds onto four minibands. We find that a small  $V_0$ is sufficient to induce CDW gaps, resulting an insulating ground state at $\nu=1/2$. The CDW-reconstructed band structure is shown in Fig \ref{fig:pheno}(b). Moreover, we find that the lowest four minibands of holes carry Chern numbers $C = 1$, $C = 0$, $ C = -1$ and $C = 1$ respectively. At half filling $\nu = 1/2$ , corresponding to completely filling the lowest two minibands, our system with $2\times 2$ CDW order has a ground-state Chern number $ C = 1$, and therefore is a QAH crystal with $\sigma_{xy}=1$.

To summarize, our main finding is a QAH crystal phase at $\nu=1/2$ in $t$MoTe$_2$ at larger twist angles than hitertho studied. This state is interesting for a number of reasons. First, it exhibits a quantized Hall conductance that is unexpected from the filling factor \cite{kol1993fractional}: $\sigma_{xy}\neq \nu C_{\rm band}$, unlike all observed QAH states. 
Second, it exhibits a $2\times 2$ bidirectional CDW order rather than the $2\times 1$ 
stripe order observed or proposed in other moir\'e materials \cite{jinStripePhasesWSe22021, li2021imaging, Polshyn_2021}. Third, as a function of twist angle, there appears to be a direct transition between CFL and QAH crystal (although the possibility of a narrow intermediate phase cannot be excluded).

Our work raises new questions for future investigation. While our study is focused on $t$MoTe$_2$ at zero displacement field, it will be interesting to explore the possibility of inducing crystal and QAH crystal states by applying a displacement field. The nature of the CFL to QAH crystal phase transition calls for theoretical understanding.   
Last but not the least, other physical realizations of  QAH- and FQAH crystals in moir\'e superlattices \cite{song2023, lu2024,  kourtis2018symmetry, daghofer2014}, as well as the theoretical possibility of QAH crystals without lattice effect \cite{dong2023,zhou2023fractional}, remain to be explored.

\begin{acknowledgements}
{\it Acknowledgments.---} We thank Xiaodong Xu for insightful discussions. 
D.N.S. was supported by the U.S. Department of
Energy, Office of Basic Energy Sciences under Grant No. DE-FG02-06ER46305. 
The work at Massachusetts Institute of Technology was supported by the
Air Force Office of Scientific Research (AFOSR) under Award No. FA9550-22-1-0432 and benefited from computing resources provided by the MIT SuperCloud and Lincoln Laboratory Supercomputing Center. A.A was supported by the Knut and Alice Wallenberg Foundation (KAW 2022.0348). E.J.B. was supported by the Swedish Research Council (Grant No. 2018-00313), Knut and Alice Wallenberg Foundation (2018.0460) and the G\"oran Gustafsson Foundation for Research in Natural Sciences and Medicine. L. F. was partly supported by the Simons Investigator Award from the Simons Foundation.   

\end{acknowledgements}

\bibliography{ref}

\begin{thebibliography}{57}%
\makeatletter
\providecommand \@ifxundefined [1]{%
 \@ifx{#1\undefined}
}%
\providecommand \@ifnum [1]{%
 \ifnum #1\expandafter \@firstoftwo
 \else \expandafter \@secondoftwo
 \fi
}%
\providecommand \@ifx [1]{%
 \ifx #1\expandafter \@firstoftwo
 \else \expandafter \@secondoftwo
 \fi
}%
\providecommand \natexlab [1]{#1}%
\providecommand \enquote  [1]{``#1''}%
\providecommand \bibnamefont  [1]{#1}%
\providecommand \bibfnamefont [1]{#1}%
\providecommand \citenamefont [1]{#1}%
\providecommand \href@noop [0]{\@secondoftwo}%
\providecommand \href [0]{\begingroup \@sanitize@url \@href}%
\providecommand \@href[1]{\@@startlink{#1}\@@href}%
\providecommand \@@href[1]{\endgroup#1\@@endlink}%
\providecommand \@sanitize@url [0]{\catcode `\\12\catcode `\$12\catcode
  `\&12\catcode `\#12\catcode `\^12\catcode `\_12\catcode `\%12\relax}%
\providecommand \@@startlink[1]{}%
\providecommand \@@endlink[0]{}%
\providecommand \url  [0]{\begingroup\@sanitize@url \@url }%
\providecommand \@url [1]{\endgroup\@href {#1}{\urlprefix }}%
\providecommand \urlprefix  [0]{URL }%
\providecommand \Eprint [0]{\href }%
\providecommand \doibase [0]{https://doi.org/}%
\providecommand \selectlanguage [0]{\@gobble}%
\providecommand \bibinfo  [0]{\@secondoftwo}%
\providecommand \bibfield  [0]{\@secondoftwo}%
\providecommand \translation [1]{[#1]}%
\providecommand \BibitemOpen [0]{}%
\providecommand \bibitemStop [0]{}%
\providecommand \bibitemNoStop [0]{.\EOS\space}%
\providecommand \EOS [0]{\spacefactor3000\relax}%
\providecommand \BibitemShut  [1]{\csname bibitem#1\endcsname}%
\let\auto@bib@innerbib\@empty
\bibitem [{\citenamefont {Andrei}\ and\ \citenamefont
  {MacDonald}(2020)}]{andrei_graphene_2020}%
  \BibitemOpen
  \bibfield  {author} {\bibinfo {author} {\bibfnamefont {E.~Y.}\ \bibnamefont
  {Andrei}}\ and\ \bibinfo {author} {\bibfnamefont {A.~H.}\ \bibnamefont
  {MacDonald}},\ }\bibfield  {title} {\bibinfo {title} {Graphene bilayers with
  a twist},\ }\href {https://doi.org/10.1038/s41563-020-00840-0} {\bibfield
  {journal} {\bibinfo  {journal} {Nature Materials}\ }\textbf {\bibinfo
  {volume} {19}},\ \bibinfo {pages} {1265} (\bibinfo {year}
  {2020})}\BibitemShut {NoStop}%
\bibitem [{\citenamefont {Mak}\ and\ \citenamefont
  {Shan}(2022)}]{mak_semiconductor_2022}%
  \BibitemOpen
  \bibfield  {author} {\bibinfo {author} {\bibfnamefont {K.~F.}\ \bibnamefont
  {Mak}}\ and\ \bibinfo {author} {\bibfnamefont {J.}~\bibnamefont {Shan}},\
  }\bibfield  {title} {\bibinfo {title} {Semiconductor moiré materials},\
  }\href {https://doi.org/10.1038/s41565-022-01165-6} {\bibfield  {journal}
  {\bibinfo  {journal} {Nature Nanotechnology}\ }\textbf {\bibinfo {volume}
  {17}},\ \bibinfo {pages} {686} (\bibinfo {year} {2022})}\BibitemShut
  {NoStop}%
\bibitem [{\citenamefont {Cao}\ \emph {et~al.}(2018)\citenamefont {Cao},
  \citenamefont {Fatemi}, \citenamefont {Fang}, \citenamefont {Watanabe},
  \citenamefont {Taniguchi}, \citenamefont {Kaxiras},\ and\ \citenamefont
  {{Jarillo-Herrero}}}]{caoUnconventionalSuperconductivityMagicangle2018}%
  \BibitemOpen
  \bibfield  {author} {\bibinfo {author} {\bibfnamefont {Y.}~\bibnamefont
  {Cao}}, \bibinfo {author} {\bibfnamefont {V.}~\bibnamefont {Fatemi}},
  \bibinfo {author} {\bibfnamefont {S.}~\bibnamefont {Fang}}, \bibinfo {author}
  {\bibfnamefont {K.}~\bibnamefont {Watanabe}}, \bibinfo {author}
  {\bibfnamefont {T.}~\bibnamefont {Taniguchi}}, \bibinfo {author}
  {\bibfnamefont {E.}~\bibnamefont {Kaxiras}},\ and\ \bibinfo {author}
  {\bibfnamefont {P.}~\bibnamefont {{Jarillo-Herrero}}},\ }\bibfield  {title}
  {\bibinfo {title} {Unconventional superconductivity in magic-angle graphene
  superlattices},\ }\href {https://doi.org/10.1038/nature26160} {\bibfield
  {journal} {\bibinfo  {journal} {Nature}\ }\textbf {\bibinfo {volume} {556}},\
  \bibinfo {pages} {43} (\bibinfo {year} {2018})}\BibitemShut {NoStop}%
\bibitem [{\citenamefont {Sharpe}\ \emph {et~al.}(2019)\citenamefont {Sharpe},
  \citenamefont {Fox}, \citenamefont {Barnard}, \citenamefont {Finney},
  \citenamefont {Watanabe}, \citenamefont {Taniguchi}, \citenamefont
  {Kastner},\ and\ \citenamefont {Goldhaber-Gordon}}]{sharpe2019emergent}%
  \BibitemOpen
  \bibfield  {author} {\bibinfo {author} {\bibfnamefont {A.~L.}\ \bibnamefont
  {Sharpe}}, \bibinfo {author} {\bibfnamefont {E.~J.}\ \bibnamefont {Fox}},
  \bibinfo {author} {\bibfnamefont {A.~W.}\ \bibnamefont {Barnard}}, \bibinfo
  {author} {\bibfnamefont {J.}~\bibnamefont {Finney}}, \bibinfo {author}
  {\bibfnamefont {K.}~\bibnamefont {Watanabe}}, \bibinfo {author}
  {\bibfnamefont {T.}~\bibnamefont {Taniguchi}}, \bibinfo {author}
  {\bibfnamefont {M.}~\bibnamefont {Kastner}},\ and\ \bibinfo {author}
  {\bibfnamefont {D.}~\bibnamefont {Goldhaber-Gordon}},\ }\bibfield  {title}
  {\bibinfo {title} {Emergent ferromagnetism near three-quarters filling in
  twisted bilayer graphene},\ }\href
  {https://www.science.org/doi/10.1126/science.aaw3780} {\bibfield  {journal}
  {\bibinfo  {journal} {Science}\ }\textbf {\bibinfo {volume} {365}},\ \bibinfo
  {pages} {605} (\bibinfo {year} {2019})}\BibitemShut {NoStop}%
\bibitem [{\citenamefont {Anderson}\ \emph {et~al.}(2023)\citenamefont
  {Anderson}, \citenamefont {Fan}, \citenamefont {Cai}, \citenamefont
  {Holtzmann}, \citenamefont {Taniguchi}, \citenamefont {Watanabe},
  \citenamefont {Xiao}, \citenamefont {Yao},\ and\ \citenamefont
  {Xu}}]{anderson2023programming}%
  \BibitemOpen
  \bibfield  {author} {\bibinfo {author} {\bibfnamefont {E.}~\bibnamefont
  {Anderson}}, \bibinfo {author} {\bibfnamefont {F.-R.}\ \bibnamefont {Fan}},
  \bibinfo {author} {\bibfnamefont {J.}~\bibnamefont {Cai}}, \bibinfo {author}
  {\bibfnamefont {W.}~\bibnamefont {Holtzmann}}, \bibinfo {author}
  {\bibfnamefont {T.}~\bibnamefont {Taniguchi}}, \bibinfo {author}
  {\bibfnamefont {K.}~\bibnamefont {Watanabe}}, \bibinfo {author}
  {\bibfnamefont {D.}~\bibnamefont {Xiao}}, \bibinfo {author} {\bibfnamefont
  {W.}~\bibnamefont {Yao}},\ and\ \bibinfo {author} {\bibfnamefont
  {X.}~\bibnamefont {Xu}},\ }\bibfield  {title} {\bibinfo {title} {Programming
  correlated magnetic states with gate-controlled moir{\'e} geometry},\ }\href
  {https://www.science.org/doi/10.1126/science.adg4268} {\bibfield  {journal}
  {\bibinfo  {journal} {Science}\ }\textbf {\bibinfo {volume} {381}},\ \bibinfo
  {pages} {325} (\bibinfo {year} {2023})}\BibitemShut {NoStop}%
\bibitem [{\citenamefont {Li}\ \emph {et~al.}(2021{\natexlab{a}})\citenamefont
  {Li}, \citenamefont {Li}, \citenamefont {Regan}, \citenamefont {Wang},
  \citenamefont {Zhao}, \citenamefont {Kahn}, \citenamefont {Yumigeta},
  \citenamefont {Blei}, \citenamefont {Taniguchi}, \citenamefont {Watanabe}
  \emph {et~al.}}]{li2021imaging}%
  \BibitemOpen
  \bibfield  {author} {\bibinfo {author} {\bibfnamefont {H.}~\bibnamefont
  {Li}}, \bibinfo {author} {\bibfnamefont {S.}~\bibnamefont {Li}}, \bibinfo
  {author} {\bibfnamefont {E.~C.}\ \bibnamefont {Regan}}, \bibinfo {author}
  {\bibfnamefont {D.}~\bibnamefont {Wang}}, \bibinfo {author} {\bibfnamefont
  {W.}~\bibnamefont {Zhao}}, \bibinfo {author} {\bibfnamefont {S.}~\bibnamefont
  {Kahn}}, \bibinfo {author} {\bibfnamefont {K.}~\bibnamefont {Yumigeta}},
  \bibinfo {author} {\bibfnamefont {M.}~\bibnamefont {Blei}}, \bibinfo {author}
  {\bibfnamefont {T.}~\bibnamefont {Taniguchi}}, \bibinfo {author}
  {\bibfnamefont {K.}~\bibnamefont {Watanabe}}, \emph {et~al.},\ }\bibfield
  {title} {\bibinfo {title} {Imaging two-dimensional generalized wigner
  crystals},\ }\href {https://doi.org/10.1038/s41586-021-03874-9} {\bibfield
  {journal} {\bibinfo  {journal} {Nature}\ }\textbf {\bibinfo {volume} {597}},\
  \bibinfo {pages} {650} (\bibinfo {year} {2021}{\natexlab{a}})}\BibitemShut
  {NoStop}%
\bibitem [{\citenamefont {Jin}\ \emph {et~al.}(2021)\citenamefont {Jin},
  \citenamefont {Tao}, \citenamefont {Li}, \citenamefont {Xu}, \citenamefont
  {Tang}, \citenamefont {Zhu}, \citenamefont {Liu}, \citenamefont {Watanabe},
  \citenamefont {Taniguchi}, \citenamefont {Hone}, \citenamefont {Fu},
  \citenamefont {Shan},\ and\ \citenamefont {Mak}}]{jinStripePhasesWSe22021}%
  \BibitemOpen
  \bibfield  {author} {\bibinfo {author} {\bibfnamefont {C.}~\bibnamefont
  {Jin}}, \bibinfo {author} {\bibfnamefont {Z.}~\bibnamefont {Tao}}, \bibinfo
  {author} {\bibfnamefont {T.}~\bibnamefont {Li}}, \bibinfo {author}
  {\bibfnamefont {Y.}~\bibnamefont {Xu}}, \bibinfo {author} {\bibfnamefont
  {Y.}~\bibnamefont {Tang}}, \bibinfo {author} {\bibfnamefont {J.}~\bibnamefont
  {Zhu}}, \bibinfo {author} {\bibfnamefont {S.}~\bibnamefont {Liu}}, \bibinfo
  {author} {\bibfnamefont {K.}~\bibnamefont {Watanabe}}, \bibinfo {author}
  {\bibfnamefont {T.}~\bibnamefont {Taniguchi}}, \bibinfo {author}
  {\bibfnamefont {J.~C.}\ \bibnamefont {Hone}}, \bibinfo {author}
  {\bibfnamefont {L.}~\bibnamefont {Fu}}, \bibinfo {author} {\bibfnamefont
  {J.}~\bibnamefont {Shan}},\ and\ \bibinfo {author} {\bibfnamefont {K.~F.}\
  \bibnamefont {Mak}},\ }\bibfield  {title} {\bibinfo {title} {Stripe phases in
  {{WSe2}}/{{WS2}} moir\'e superlattices},\ }\href
  {https://doi.org/10.1038/s41563-021-00959-8} {\bibfield  {journal} {\bibinfo
  {journal} {Nat. Mater.}\ }\textbf {\bibinfo {volume} {20}},\ \bibinfo {pages}
  {940} (\bibinfo {year} {2021})}\BibitemShut {NoStop}%
\bibitem [{\citenamefont {Dean}\ \emph {et~al.}(2013)\citenamefont {Dean},
  \citenamefont {Wang}, \citenamefont {Maher}, \citenamefont {Forsythe},
  \citenamefont {Ghahari}, \citenamefont {Gao}, \citenamefont {Katoch},
  \citenamefont {Ishigami}, \citenamefont {Moon}, \citenamefont {Koshino},
  \citenamefont {Taniguchi}, \citenamefont {Watanabe}, \citenamefont {Shepard},
  \citenamefont {Hone},\ and\ \citenamefont {Kim}}]{dean_hofstadters_2013}%
  \BibitemOpen
  \bibfield  {author} {\bibinfo {author} {\bibfnamefont {C.~R.}\ \bibnamefont
  {Dean}}, \bibinfo {author} {\bibfnamefont {L.}~\bibnamefont {Wang}}, \bibinfo
  {author} {\bibfnamefont {P.}~\bibnamefont {Maher}}, \bibinfo {author}
  {\bibfnamefont {C.}~\bibnamefont {Forsythe}}, \bibinfo {author}
  {\bibfnamefont {F.}~\bibnamefont {Ghahari}}, \bibinfo {author} {\bibfnamefont
  {Y.}~\bibnamefont {Gao}}, \bibinfo {author} {\bibfnamefont {J.}~\bibnamefont
  {Katoch}}, \bibinfo {author} {\bibfnamefont {M.}~\bibnamefont {Ishigami}},
  \bibinfo {author} {\bibfnamefont {P.}~\bibnamefont {Moon}}, \bibinfo {author}
  {\bibfnamefont {M.}~\bibnamefont {Koshino}}, \bibinfo {author} {\bibfnamefont
  {T.}~\bibnamefont {Taniguchi}}, \bibinfo {author} {\bibfnamefont
  {K.}~\bibnamefont {Watanabe}}, \bibinfo {author} {\bibfnamefont {K.~L.}\
  \bibnamefont {Shepard}}, \bibinfo {author} {\bibfnamefont {J.}~\bibnamefont
  {Hone}},\ and\ \bibinfo {author} {\bibfnamefont {P.}~\bibnamefont {Kim}},\
  }\bibfield  {title} {\bibinfo {title} {Hofstadter’s butterfly and the
  fractal quantum {Hall} effect in moiré superlattices},\ }\href
  {https://doi.org/10.1038/nature12186} {\bibfield  {journal} {\bibinfo
  {journal} {Nature}\ }\textbf {\bibinfo {volume} {497}},\ \bibinfo {pages}
  {598} (\bibinfo {year} {2013})}\BibitemShut {NoStop}%
\bibitem [{\citenamefont {Spanton}\ \emph {et~al.}(2018)\citenamefont
  {Spanton}, \citenamefont {Zibrov}, \citenamefont {Zhou}, \citenamefont
  {Taniguchi}, \citenamefont {Watanabe}, \citenamefont {Zaletel},\ and\
  \citenamefont {Young}}]{spanton2018observation}%
  \BibitemOpen
  \bibfield  {author} {\bibinfo {author} {\bibfnamefont {E.~M.}\ \bibnamefont
  {Spanton}}, \bibinfo {author} {\bibfnamefont {A.~A.}\ \bibnamefont {Zibrov}},
  \bibinfo {author} {\bibfnamefont {H.}~\bibnamefont {Zhou}}, \bibinfo {author}
  {\bibfnamefont {T.}~\bibnamefont {Taniguchi}}, \bibinfo {author}
  {\bibfnamefont {K.}~\bibnamefont {Watanabe}}, \bibinfo {author}
  {\bibfnamefont {M.~P.}\ \bibnamefont {Zaletel}},\ and\ \bibinfo {author}
  {\bibfnamefont {A.~F.}\ \bibnamefont {Young}},\ }\bibfield  {title} {\bibinfo
  {title} {Observation of fractional chern insulators in a van der waals
  heterostructure},\ }\href
  {https://www.science.org/doi/10.1126/science.aan8458} {\bibfield  {journal}
  {\bibinfo  {journal} {Science}\ }\textbf {\bibinfo {volume} {360}},\ \bibinfo
  {pages} {62} (\bibinfo {year} {2018})}\BibitemShut {NoStop}%
\bibitem [{\citenamefont {Kang}\ \emph
  {et~al.}(2024{\natexlab{a}})\citenamefont {Kang}, \citenamefont {Shen},
  \citenamefont {Qiu}, \citenamefont {Watanabe}, \citenamefont {Taniguchi},
  \citenamefont {Shan},\ and\ \citenamefont {Mak}}]{kang_observation_2024}%
  \BibitemOpen
  \bibfield  {author} {\bibinfo {author} {\bibfnamefont {K.}~\bibnamefont
  {Kang}}, \bibinfo {author} {\bibfnamefont {B.}~\bibnamefont {Shen}}, \bibinfo
  {author} {\bibfnamefont {Y.}~\bibnamefont {Qiu}}, \bibinfo {author}
  {\bibfnamefont {K.}~\bibnamefont {Watanabe}}, \bibinfo {author}
  {\bibfnamefont {T.}~\bibnamefont {Taniguchi}}, \bibinfo {author}
  {\bibfnamefont {J.}~\bibnamefont {Shan}},\ and\ \bibinfo {author}
  {\bibfnamefont {K.~F.}\ \bibnamefont {Mak}},\ }\bibfield  {title} {\bibinfo
  {title} {Observation of the fractional quantum spin {Hall} effect in moir\'e
  {MoTe2}},\ }\href {http://arxiv.org/abs/2402.03294} {\bibfield  {journal}
  {\bibinfo  {journal} {arXiv preprint arXiv:2402.03294}\ } (\bibinfo {year}
  {2024}{\natexlab{a}})}\BibitemShut {NoStop}%
\bibitem [{\citenamefont {Kang}\ \emph
  {et~al.}(2024{\natexlab{b}})\citenamefont {Kang}, \citenamefont {Qiu},
  \citenamefont {Watanabe}, \citenamefont {Taniguchi}, \citenamefont {Shan},\
  and\ \citenamefont {Mak}}]{kang2024observation2}%
  \BibitemOpen
  \bibfield  {author} {\bibinfo {author} {\bibfnamefont {K.}~\bibnamefont
  {Kang}}, \bibinfo {author} {\bibfnamefont {Y.}~\bibnamefont {Qiu}}, \bibinfo
  {author} {\bibfnamefont {K.}~\bibnamefont {Watanabe}}, \bibinfo {author}
  {\bibfnamefont {T.}~\bibnamefont {Taniguchi}}, \bibinfo {author}
  {\bibfnamefont {J.}~\bibnamefont {Shan}},\ and\ \bibinfo {author}
  {\bibfnamefont {K.~F.}\ \bibnamefont {Mak}},\ }\bibfield  {title} {\bibinfo
  {title} {Observation of the double quantum spin hall phase in
  moir$\backslash$'e wse2},\ }\href {https://arxiv.org/abs/2402.04196}
  {\bibfield  {journal} {\bibinfo  {journal} {arXiv preprint arXiv:2402.04196}\
  } (\bibinfo {year} {2024}{\natexlab{b}})}\BibitemShut {NoStop}%
\bibitem [{\citenamefont {Serlin}\ \emph {et~al.}(2020)\citenamefont {Serlin},
  \citenamefont {Tschirhart}, \citenamefont {Polshyn}, \citenamefont {Zhang},
  \citenamefont {Zhu}, \citenamefont {Watanabe}, \citenamefont {Taniguchi},
  \citenamefont {Balents},\ and\ \citenamefont {Young}}]{serlin2020intrinsic}%
  \BibitemOpen
  \bibfield  {author} {\bibinfo {author} {\bibfnamefont {M.}~\bibnamefont
  {Serlin}}, \bibinfo {author} {\bibfnamefont {C.}~\bibnamefont {Tschirhart}},
  \bibinfo {author} {\bibfnamefont {H.}~\bibnamefont {Polshyn}}, \bibinfo
  {author} {\bibfnamefont {Y.}~\bibnamefont {Zhang}}, \bibinfo {author}
  {\bibfnamefont {J.}~\bibnamefont {Zhu}}, \bibinfo {author} {\bibfnamefont
  {K.}~\bibnamefont {Watanabe}}, \bibinfo {author} {\bibfnamefont
  {T.}~\bibnamefont {Taniguchi}}, \bibinfo {author} {\bibfnamefont
  {L.}~\bibnamefont {Balents}},\ and\ \bibinfo {author} {\bibfnamefont
  {A.}~\bibnamefont {Young}},\ }\bibfield  {title} {\bibinfo {title} {Intrinsic
  quantized anomalous hall effect in a moir{\'e} heterostructure},\ }\href
  {https://doi.org/10.1126/science.aay5533} {\bibfield  {journal} {\bibinfo
  {journal} {Science}\ }\textbf {\bibinfo {volume} {367}},\ \bibinfo {pages}
  {900} (\bibinfo {year} {2020})}\BibitemShut {NoStop}%
\bibitem [{\citenamefont {Li}\ \emph {et~al.}(2021{\natexlab{b}})\citenamefont
  {Li}, \citenamefont {Jiang}, \citenamefont {Shen}, \citenamefont {Zhang},
  \citenamefont {Li}, \citenamefont {Tao}, \citenamefont {Devakul},
  \citenamefont {Watanabe}, \citenamefont {Taniguchi}, \citenamefont {Fu} \emph
  {et~al.}}]{li2021quantum}%
  \BibitemOpen
  \bibfield  {author} {\bibinfo {author} {\bibfnamefont {T.}~\bibnamefont
  {Li}}, \bibinfo {author} {\bibfnamefont {S.}~\bibnamefont {Jiang}}, \bibinfo
  {author} {\bibfnamefont {B.}~\bibnamefont {Shen}}, \bibinfo {author}
  {\bibfnamefont {Y.}~\bibnamefont {Zhang}}, \bibinfo {author} {\bibfnamefont
  {L.}~\bibnamefont {Li}}, \bibinfo {author} {\bibfnamefont {Z.}~\bibnamefont
  {Tao}}, \bibinfo {author} {\bibfnamefont {T.}~\bibnamefont {Devakul}},
  \bibinfo {author} {\bibfnamefont {K.}~\bibnamefont {Watanabe}}, \bibinfo
  {author} {\bibfnamefont {T.}~\bibnamefont {Taniguchi}}, \bibinfo {author}
  {\bibfnamefont {L.}~\bibnamefont {Fu}}, \emph {et~al.},\ }\bibfield  {title}
  {\bibinfo {title} {Quantum anomalous hall effect from intertwined moir{\'e}
  bands},\ }\href {https://doi.org/10.1038/s41586-021-04171-1} {\bibfield
  {journal} {\bibinfo  {journal} {Nature}\ }\textbf {\bibinfo {volume} {600}},\
  \bibinfo {pages} {641} (\bibinfo {year} {2021}{\natexlab{b}})}\BibitemShut
  {NoStop}%
\bibitem [{\citenamefont {Foutty}\ \emph {et~al.}(2024)\citenamefont {Foutty},
  \citenamefont {Kometter}, \citenamefont {Devakul}, \citenamefont {Reddy},
  \citenamefont {Watanabe}, \citenamefont {Taniguchi}, \citenamefont {Fu},\
  and\ \citenamefont {Feldman}}]{foutty2023mapping}%
  \BibitemOpen
  \bibfield  {author} {\bibinfo {author} {\bibfnamefont {B.~A.}\ \bibnamefont
  {Foutty}}, \bibinfo {author} {\bibfnamefont {C.~R.}\ \bibnamefont
  {Kometter}}, \bibinfo {author} {\bibfnamefont {T.}~\bibnamefont {Devakul}},
  \bibinfo {author} {\bibfnamefont {A.~P.}\ \bibnamefont {Reddy}}, \bibinfo
  {author} {\bibfnamefont {K.}~\bibnamefont {Watanabe}}, \bibinfo {author}
  {\bibfnamefont {T.}~\bibnamefont {Taniguchi}}, \bibinfo {author}
  {\bibfnamefont {L.}~\bibnamefont {Fu}},\ and\ \bibinfo {author}
  {\bibfnamefont {B.~E.}\ \bibnamefont {Feldman}},\ }\bibfield  {title}
  {\bibinfo {title} {Mapping twist-tuned multi-band topology in bilayer
  wse$_2$},\ }\href {https://www.science.org/doi/10.1126/science.adi4728}
  {\bibfield  {journal} {\bibinfo  {journal} {Science}\ }\textbf {\bibinfo
  {volume} {384}},\ \bibinfo {pages} {343} (\bibinfo {year}
  {2024})}\BibitemShut {NoStop}%
\bibitem [{\citenamefont {Park}\ \emph {et~al.}(2023)\citenamefont {Park},
  \citenamefont {Cai}, \citenamefont {Anderson}, \citenamefont {Zhang},
  \citenamefont {Zhu}, \citenamefont {Liu}, \citenamefont {Wang}, \citenamefont
  {Holtzmann}, \citenamefont {Hu}, \citenamefont {Liu}, \citenamefont
  {Taniguchi}, \citenamefont {Watanabe}, \citenamefont {Chu}, \citenamefont
  {Cao}, \citenamefont {Fu}, \citenamefont {Yao}, \citenamefont {Chang},
  \citenamefont {Cobden}, \citenamefont {Xiao},\ and\ \citenamefont
  {Xu}}]{park2023observation}%
  \BibitemOpen
  \bibfield  {author} {\bibinfo {author} {\bibfnamefont {H.}~\bibnamefont
  {Park}}, \bibinfo {author} {\bibfnamefont {J.}~\bibnamefont {Cai}}, \bibinfo
  {author} {\bibfnamefont {E.}~\bibnamefont {Anderson}}, \bibinfo {author}
  {\bibfnamefont {Y.}~\bibnamefont {Zhang}}, \bibinfo {author} {\bibfnamefont
  {J.}~\bibnamefont {Zhu}}, \bibinfo {author} {\bibfnamefont {X.}~\bibnamefont
  {Liu}}, \bibinfo {author} {\bibfnamefont {C.}~\bibnamefont {Wang}}, \bibinfo
  {author} {\bibfnamefont {W.}~\bibnamefont {Holtzmann}}, \bibinfo {author}
  {\bibfnamefont {C.}~\bibnamefont {Hu}}, \bibinfo {author} {\bibfnamefont
  {Z.}~\bibnamefont {Liu}}, \bibinfo {author} {\bibfnamefont {T.}~\bibnamefont
  {Taniguchi}}, \bibinfo {author} {\bibfnamefont {K.}~\bibnamefont {Watanabe}},
  \bibinfo {author} {\bibfnamefont {J.-H.}\ \bibnamefont {Chu}}, \bibinfo
  {author} {\bibfnamefont {T.}~\bibnamefont {Cao}}, \bibinfo {author}
  {\bibfnamefont {L.}~\bibnamefont {Fu}}, \bibinfo {author} {\bibfnamefont
  {W.}~\bibnamefont {Yao}}, \bibinfo {author} {\bibfnamefont {C.-Z.}\
  \bibnamefont {Chang}}, \bibinfo {author} {\bibfnamefont {D.}~\bibnamefont
  {Cobden}}, \bibinfo {author} {\bibfnamefont {D.}~\bibnamefont {Xiao}},\ and\
  \bibinfo {author} {\bibfnamefont {X.}~\bibnamefont {Xu}},\ }\bibfield
  {title} {\bibinfo {title} {Observation of fractionally quantized anomalous
  hall effect},\ }\href {https://doi.org/10.1038/s41586-023-06536-0} {\bibfield
   {journal} {\bibinfo  {journal} {Nature}\ }\textbf {\bibinfo {volume}
  {622}},\ \bibinfo {pages} {74} (\bibinfo {year} {2023})}\BibitemShut
  {NoStop}%
\bibitem [{\citenamefont {Xu}\ \emph {et~al.}(2023)\citenamefont {Xu},
  \citenamefont {Sun}, \citenamefont {Jia}, \citenamefont {Liu}, \citenamefont
  {Xu}, \citenamefont {Li}, \citenamefont {Gu}, \citenamefont {Watanabe},
  \citenamefont {Taniguchi}, \citenamefont {Tong}, \citenamefont {Jia},
  \citenamefont {Shi}, \citenamefont {Jiang}, \citenamefont {Zhang},
  \citenamefont {Liu},\ and\ \citenamefont {Li}}]{xu_observation_2023}%
  \BibitemOpen
  \bibfield  {author} {\bibinfo {author} {\bibfnamefont {F.}~\bibnamefont
  {Xu}}, \bibinfo {author} {\bibfnamefont {Z.}~\bibnamefont {Sun}}, \bibinfo
  {author} {\bibfnamefont {T.}~\bibnamefont {Jia}}, \bibinfo {author}
  {\bibfnamefont {C.}~\bibnamefont {Liu}}, \bibinfo {author} {\bibfnamefont
  {C.}~\bibnamefont {Xu}}, \bibinfo {author} {\bibfnamefont {C.}~\bibnamefont
  {Li}}, \bibinfo {author} {\bibfnamefont {Y.}~\bibnamefont {Gu}}, \bibinfo
  {author} {\bibfnamefont {K.}~\bibnamefont {Watanabe}}, \bibinfo {author}
  {\bibfnamefont {T.}~\bibnamefont {Taniguchi}}, \bibinfo {author}
  {\bibfnamefont {B.}~\bibnamefont {Tong}}, \bibinfo {author} {\bibfnamefont
  {J.}~\bibnamefont {Jia}}, \bibinfo {author} {\bibfnamefont {Z.}~\bibnamefont
  {Shi}}, \bibinfo {author} {\bibfnamefont {S.}~\bibnamefont {Jiang}}, \bibinfo
  {author} {\bibfnamefont {Y.}~\bibnamefont {Zhang}}, \bibinfo {author}
  {\bibfnamefont {X.}~\bibnamefont {Liu}},\ and\ \bibinfo {author}
  {\bibfnamefont {T.}~\bibnamefont {Li}},\ }\bibfield  {title} {\bibinfo
  {title} {Observation of {Integer} and {Fractional} {Quantum} {Anomalous}
  {Hall} {Effects} in {Twisted} {Bilayer}
  \$\{{\textbackslash}mathrm\{{MoTe}\}\}\_\{2\}\$},\ }\href
  {https://doi.org/10.1103/PhysRevX.13.031037} {\bibfield  {journal} {\bibinfo
  {journal} {Physical Review X}\ }\textbf {\bibinfo {volume} {13}},\ \bibinfo
  {pages} {031037} (\bibinfo {year} {2023})}\BibitemShut {NoStop}%
\bibitem [{\citenamefont {Cai}\ \emph {et~al.}(2023)\citenamefont {Cai},
  \citenamefont {Anderson}, \citenamefont {Wang}, \citenamefont {Zhang},
  \citenamefont {Liu}, \citenamefont {Holtzmann}, \citenamefont {Zhang},
  \citenamefont {Fan}, \citenamefont {Taniguchi}, \citenamefont {Watanabe},
  \citenamefont {Ran}, \citenamefont {Cao}, \citenamefont {Fu}, \citenamefont
  {Xiao}, \citenamefont {Yao},\ and\ \citenamefont {Xu}}]{cai2023signatures}%
  \BibitemOpen
  \bibfield  {author} {\bibinfo {author} {\bibfnamefont {J.}~\bibnamefont
  {Cai}}, \bibinfo {author} {\bibfnamefont {E.}~\bibnamefont {Anderson}},
  \bibinfo {author} {\bibfnamefont {C.}~\bibnamefont {Wang}}, \bibinfo {author}
  {\bibfnamefont {X.}~\bibnamefont {Zhang}}, \bibinfo {author} {\bibfnamefont
  {X.}~\bibnamefont {Liu}}, \bibinfo {author} {\bibfnamefont {W.}~\bibnamefont
  {Holtzmann}}, \bibinfo {author} {\bibfnamefont {Y.}~\bibnamefont {Zhang}},
  \bibinfo {author} {\bibfnamefont {F.}~\bibnamefont {Fan}}, \bibinfo {author}
  {\bibfnamefont {T.}~\bibnamefont {Taniguchi}}, \bibinfo {author}
  {\bibfnamefont {K.}~\bibnamefont {Watanabe}}, \bibinfo {author}
  {\bibfnamefont {Y.}~\bibnamefont {Ran}}, \bibinfo {author} {\bibfnamefont
  {T.}~\bibnamefont {Cao}}, \bibinfo {author} {\bibfnamefont {L.}~\bibnamefont
  {Fu}}, \bibinfo {author} {\bibfnamefont {D.}~\bibnamefont {Xiao}}, \bibinfo
  {author} {\bibfnamefont {W.}~\bibnamefont {Yao}},\ and\ \bibinfo {author}
  {\bibfnamefont {X.}~\bibnamefont {Xu}},\ }\bibfield  {title} {\bibinfo
  {title} {Signatures of fractional quantum anomalous hall states in twisted
  mote2},\ }\href {https://doi.org/10.1038/s41586-023-06289-w} {\bibfield
  {journal} {\bibinfo  {journal} {Nature}\ }\textbf {\bibinfo {volume} {622}},\
  \bibinfo {pages} {63} (\bibinfo {year} {2023})}\BibitemShut {NoStop}%
\bibitem [{\citenamefont {Zeng}\ \emph {et~al.}(2023)\citenamefont {Zeng},
  \citenamefont {Xia}, \citenamefont {Kang}, \citenamefont {Zhu}, \citenamefont
  {Kn{\"u}ppel}, \citenamefont {Vaswani}, \citenamefont {Watanabe},
  \citenamefont {Taniguchi}, \citenamefont {Mak},\ and\ \citenamefont
  {Shan}}]{zeng2023thermodynamic}%
  \BibitemOpen
  \bibfield  {author} {\bibinfo {author} {\bibfnamefont {Y.}~\bibnamefont
  {Zeng}}, \bibinfo {author} {\bibfnamefont {Z.}~\bibnamefont {Xia}}, \bibinfo
  {author} {\bibfnamefont {K.}~\bibnamefont {Kang}}, \bibinfo {author}
  {\bibfnamefont {J.}~\bibnamefont {Zhu}}, \bibinfo {author} {\bibfnamefont
  {P.}~\bibnamefont {Kn{\"u}ppel}}, \bibinfo {author} {\bibfnamefont
  {C.}~\bibnamefont {Vaswani}}, \bibinfo {author} {\bibfnamefont
  {K.}~\bibnamefont {Watanabe}}, \bibinfo {author} {\bibfnamefont
  {T.}~\bibnamefont {Taniguchi}}, \bibinfo {author} {\bibfnamefont {K.~F.}\
  \bibnamefont {Mak}},\ and\ \bibinfo {author} {\bibfnamefont {J.}~\bibnamefont
  {Shan}},\ }\bibfield  {title} {\bibinfo {title} {Thermodynamic evidence of
  fractional chern insulator in moir{\'e} mote2},\ }\href
  {https://www.nature.com/articles/s41586-023-06452-3} {\bibfield  {journal}
  {\bibinfo  {journal} {Nature}\ }\textbf {\bibinfo {volume} {622}},\ \bibinfo
  {pages} {69} (\bibinfo {year} {2023})}\BibitemShut {NoStop}%
\bibitem [{\citenamefont {Lu}\ \emph {et~al.}(2024)\citenamefont {Lu},
  \citenamefont {Han}, \citenamefont {Yao}, \citenamefont {Reddy},
  \citenamefont {Yang}, \citenamefont {Seo}, \citenamefont {Watanabe},
  \citenamefont {Taniguchi}, \citenamefont {Fu},\ and\ \citenamefont
  {Ju}}]{lu_fractional_2024}%
  \BibitemOpen
  \bibfield  {author} {\bibinfo {author} {\bibfnamefont {Z.}~\bibnamefont
  {Lu}}, \bibinfo {author} {\bibfnamefont {T.}~\bibnamefont {Han}}, \bibinfo
  {author} {\bibfnamefont {Y.}~\bibnamefont {Yao}}, \bibinfo {author}
  {\bibfnamefont {A.~P.}\ \bibnamefont {Reddy}}, \bibinfo {author}
  {\bibfnamefont {J.}~\bibnamefont {Yang}}, \bibinfo {author} {\bibfnamefont
  {J.}~\bibnamefont {Seo}}, \bibinfo {author} {\bibfnamefont {K.}~\bibnamefont
  {Watanabe}}, \bibinfo {author} {\bibfnamefont {T.}~\bibnamefont {Taniguchi}},
  \bibinfo {author} {\bibfnamefont {L.}~\bibnamefont {Fu}},\ and\ \bibinfo
  {author} {\bibfnamefont {L.}~\bibnamefont {Ju}},\ }\bibfield  {title}
  {\bibinfo {title} {Fractional quantum anomalous {Hall} effect in multilayer
  graphene},\ }\href {https://doi.org/10.1038/s41586-023-07010-7} {\bibfield
  {journal} {\bibinfo  {journal} {Nature}\ }\textbf {\bibinfo {volume} {626}},\
  \bibinfo {pages} {759} (\bibinfo {year} {2024})}\BibitemShut {NoStop}%
\bibitem [{\citenamefont {Devakul}\ \emph {et~al.}(2021)\citenamefont
  {Devakul}, \citenamefont {Cr{\'e}pel}, \citenamefont {Zhang},\ and\
  \citenamefont {Fu}}]{devakul2021magic}%
  \BibitemOpen
  \bibfield  {author} {\bibinfo {author} {\bibfnamefont {T.}~\bibnamefont
  {Devakul}}, \bibinfo {author} {\bibfnamefont {V.}~\bibnamefont {Cr{\'e}pel}},
  \bibinfo {author} {\bibfnamefont {Y.}~\bibnamefont {Zhang}},\ and\ \bibinfo
  {author} {\bibfnamefont {L.}~\bibnamefont {Fu}},\ }\bibfield  {title}
  {\bibinfo {title} {Magic in twisted transition metal dichalcogenide
  bilayers},\ }\href {https://doi.org/10.1038/s41467-021-27042-9} {\bibfield
  {journal} {\bibinfo  {journal} {Nature communications}\ }\textbf {\bibinfo
  {volume} {12}},\ \bibinfo {pages} {6730} (\bibinfo {year}
  {2021})}\BibitemShut {NoStop}%
\bibitem [{\citenamefont {Li}\ \emph {et~al.}(2021{\natexlab{c}})\citenamefont
  {Li}, \citenamefont {Kumar}, \citenamefont {Sun},\ and\ \citenamefont
  {Lin}}]{li2021spontaneous}%
  \BibitemOpen
  \bibfield  {author} {\bibinfo {author} {\bibfnamefont {H.}~\bibnamefont
  {Li}}, \bibinfo {author} {\bibfnamefont {U.}~\bibnamefont {Kumar}}, \bibinfo
  {author} {\bibfnamefont {K.}~\bibnamefont {Sun}},\ and\ \bibinfo {author}
  {\bibfnamefont {S.-Z.}\ \bibnamefont {Lin}},\ }\bibfield  {title} {\bibinfo
  {title} {Spontaneous fractional chern insulators in transition metal
  dichalcogenide moir{\'e} superlattices},\ }\href
  {https://doi.org/10.1103/PhysRevResearch.3.L032070} {\bibfield  {journal}
  {\bibinfo  {journal} {Physical Review Research}\ }\textbf {\bibinfo {volume}
  {3}},\ \bibinfo {pages} {L032070} (\bibinfo {year}
  {2021}{\natexlab{c}})}\BibitemShut {NoStop}%
\bibitem [{\citenamefont {Cr{\'e}pel}\ and\ \citenamefont
  {Fu}(2023)}]{crepel2023anomalous}%
  \BibitemOpen
  \bibfield  {author} {\bibinfo {author} {\bibfnamefont {V.}~\bibnamefont
  {Cr{\'e}pel}}\ and\ \bibinfo {author} {\bibfnamefont {L.}~\bibnamefont
  {Fu}},\ }\bibfield  {title} {\bibinfo {title} {Anomalous hall metal and
  fractional chern insulator in twisted transition metal dichalcogenides},\
  }\href {https://doi.org/10.1103/PhysRevB.107.L201109} {\bibfield  {journal}
  {\bibinfo  {journal} {Physical Review B}\ }\textbf {\bibinfo {volume}
  {107}},\ \bibinfo {pages} {L201109} (\bibinfo {year} {2023})}\BibitemShut
  {NoStop}%
\bibitem [{\citenamefont {Abouelkomsan}\ \emph {et~al.}(2020)\citenamefont
  {Abouelkomsan}, \citenamefont {Liu},\ and\ \citenamefont
  {Bergholtz}}]{abouelkomsan2020particle}%
  \BibitemOpen
  \bibfield  {author} {\bibinfo {author} {\bibfnamefont {A.}~\bibnamefont
  {Abouelkomsan}}, \bibinfo {author} {\bibfnamefont {Z.}~\bibnamefont {Liu}},\
  and\ \bibinfo {author} {\bibfnamefont {E.~J.}\ \bibnamefont {Bergholtz}},\
  }\bibfield  {title} {\bibinfo {title} {Particle-hole duality, emergent fermi
  liquids, and fractional chern insulators in moir{\'e} flatbands},\ }\href
  {https://doi.org/10.1103/PhysRevLett.124.106803} {\bibfield  {journal}
  {\bibinfo  {journal} {Physical Review Letters}\ }\textbf {\bibinfo {volume}
  {124}},\ \bibinfo {pages} {106803} (\bibinfo {year} {2020})}\BibitemShut
  {NoStop}%
\bibitem [{\citenamefont {Repellin}\ and\ \citenamefont
  {Senthil}(2020)}]{repellin2020chern}%
  \BibitemOpen
  \bibfield  {author} {\bibinfo {author} {\bibfnamefont {C.}~\bibnamefont
  {Repellin}}\ and\ \bibinfo {author} {\bibfnamefont {T.}~\bibnamefont
  {Senthil}},\ }\bibfield  {title} {\bibinfo {title} {Chern bands of twisted
  bilayer graphene: {{Fractional Chern}} insulators and spin phase
  transition},\ }\href {https://doi.org/10.1103/PhysRevResearch.2.023238}
  {\bibfield  {journal} {\bibinfo  {journal} {Physical Review Research}\
  }\textbf {\bibinfo {volume} {2}},\ \bibinfo {pages} {023238} (\bibinfo {year}
  {2020})}\BibitemShut {NoStop}%
\bibitem [{\citenamefont {Wu}\ \emph {et~al.}(2019)\citenamefont {Wu},
  \citenamefont {Lovorn}, \citenamefont {Tutuc}, \citenamefont {Martin},\ and\
  \citenamefont {MacDonald}}]{wu2019topological}%
  \BibitemOpen
  \bibfield  {author} {\bibinfo {author} {\bibfnamefont {F.}~\bibnamefont
  {Wu}}, \bibinfo {author} {\bibfnamefont {T.}~\bibnamefont {Lovorn}}, \bibinfo
  {author} {\bibfnamefont {E.}~\bibnamefont {Tutuc}}, \bibinfo {author}
  {\bibfnamefont {I.}~\bibnamefont {Martin}},\ and\ \bibinfo {author}
  {\bibfnamefont {A.}~\bibnamefont {MacDonald}},\ }\bibfield  {title} {\bibinfo
  {title} {Topological insulators in twisted transition metal dichalcogenide
  homobilayers},\ }\href {https://doi.org/10.1103/PhysRevLett.122.086402}
  {\bibfield  {journal} {\bibinfo  {journal} {Physical review letters}\
  }\textbf {\bibinfo {volume} {122}},\ \bibinfo {pages} {086402} (\bibinfo
  {year} {2019})}\BibitemShut {NoStop}%
\bibitem [{\citenamefont {Neupert}\ \emph {et~al.}(2011)\citenamefont
  {Neupert}, \citenamefont {Santos}, \citenamefont {Chamon},\ and\
  \citenamefont {Mudry}}]{neupert2011fractional}%
  \BibitemOpen
  \bibfield  {author} {\bibinfo {author} {\bibfnamefont {T.}~\bibnamefont
  {Neupert}}, \bibinfo {author} {\bibfnamefont {L.}~\bibnamefont {Santos}},
  \bibinfo {author} {\bibfnamefont {C.}~\bibnamefont {Chamon}},\ and\ \bibinfo
  {author} {\bibfnamefont {C.}~\bibnamefont {Mudry}},\ }\bibfield  {title}
  {\bibinfo {title} {Fractional quantum hall states at zero magnetic field},\
  }\href {https://doi.org/10.1103/PhysRevLett.106.236804} {\bibfield  {journal}
  {\bibinfo  {journal} {Physical review letters}\ }\textbf {\bibinfo {volume}
  {106}},\ \bibinfo {pages} {236804} (\bibinfo {year} {2011})}\BibitemShut
  {NoStop}%
\bibitem [{\citenamefont {Sheng}\ \emph {et~al.}(2011)\citenamefont {Sheng},
  \citenamefont {Gu}, \citenamefont {Sun},\ and\ \citenamefont
  {Sheng}}]{sheng2011fractional}%
  \BibitemOpen
  \bibfield  {author} {\bibinfo {author} {\bibfnamefont {D.}~\bibnamefont
  {Sheng}}, \bibinfo {author} {\bibfnamefont {Z.-C.}\ \bibnamefont {Gu}},
  \bibinfo {author} {\bibfnamefont {K.}~\bibnamefont {Sun}},\ and\ \bibinfo
  {author} {\bibfnamefont {L.}~\bibnamefont {Sheng}},\ }\bibfield  {title}
  {\bibinfo {title} {Fractional quantum hall effect in the absence of landau
  levels},\ }\href {https://doi.org/10.1038/ncomms1380} {\bibfield  {journal}
  {\bibinfo  {journal} {Nature communications}\ }\textbf {\bibinfo {volume}
  {2}},\ \bibinfo {pages} {389} (\bibinfo {year} {2011})}\BibitemShut {NoStop}%
\bibitem [{\citenamefont {Regnault}\ and\ \citenamefont
  {Bernevig}(2011)}]{regnault2011fractional}%
  \BibitemOpen
  \bibfield  {author} {\bibinfo {author} {\bibfnamefont {N.}~\bibnamefont
  {Regnault}}\ and\ \bibinfo {author} {\bibfnamefont {B.~A.}\ \bibnamefont
  {Bernevig}},\ }\bibfield  {title} {\bibinfo {title} {Fractional chern
  insulator},\ }\href
  {https://journals.aps.org/prx/abstract/10.1103/PhysRevX.1.021014} {\bibfield
  {journal} {\bibinfo  {journal} {Physical Review X}\ }\textbf {\bibinfo
  {volume} {1}},\ \bibinfo {pages} {021014} (\bibinfo {year}
  {2011})}\BibitemShut {NoStop}%
\bibitem [{\citenamefont {Liu}\ and\ \citenamefont
  {Bergholtz}(2024)}]{LIU2024515}%
  \BibitemOpen
  \bibfield  {author} {\bibinfo {author} {\bibfnamefont {Z.}~\bibnamefont
  {Liu}}\ and\ \bibinfo {author} {\bibfnamefont {E.~J.}\ \bibnamefont
  {Bergholtz}},\ }\bibfield  {title} {\bibinfo {title} {Recent developments in
  fractional chern insulators},\ }in\ \href
  {https://doi.org/https://doi.org/10.1016/B978-0-323-90800-9.00136-0} {\emph
  {\bibinfo {booktitle} {Encyclopedia of Condensed Matter Physics (Second
  Edition)}}},\ \bibinfo {editor} {edited by\ \bibinfo {editor} {\bibfnamefont
  {T.}~\bibnamefont {Chakraborty}}}\ (\bibinfo  {publisher} {Academic Press},\
  \bibinfo {address} {Oxford},\ \bibinfo {year} {2024})\ \bibinfo {edition}
  {second edition}\ ed.,\ pp.\ \bibinfo {pages} {515--538}\BibitemShut
  {NoStop}%
\bibitem [{\citenamefont {Song}\ \emph {et~al.}(2024)\citenamefont {Song},
  \citenamefont {Jian}, \citenamefont {Fu},\ and\ \citenamefont
  {Xu}}]{song2023}%
  \BibitemOpen
  \bibfield  {author} {\bibinfo {author} {\bibfnamefont {X.-Y.}\ \bibnamefont
  {Song}}, \bibinfo {author} {\bibfnamefont {C.-M.}\ \bibnamefont {Jian}},
  \bibinfo {author} {\bibfnamefont {L.}~\bibnamefont {Fu}},\ and\ \bibinfo
  {author} {\bibfnamefont {C.}~\bibnamefont {Xu}},\ }\bibfield  {title}
  {\bibinfo {title} {Intertwined fractional quantum anomalous hall states and
  charge density waves},\ }\href {https://doi.org/10.1103/PhysRevB.109.115116}
  {\bibfield  {journal} {\bibinfo  {journal} {Phys. Rev. B}\ }\textbf {\bibinfo
  {volume} {109}},\ \bibinfo {pages} {115116} (\bibinfo {year}
  {2024})}\BibitemShut {NoStop}%
\bibitem [{\citenamefont {Halperin}\ \emph {et~al.}(1986)\citenamefont
  {Halperin}, \citenamefont {Te\ifmmode \check{s}\else
  \v{s}\fi{}anovi\ifmmode~\acute{c}\else \'{c}\fi{}},\ and\ \citenamefont
  {Axel}}]{halperin1986}%
  \BibitemOpen
  \bibfield  {author} {\bibinfo {author} {\bibfnamefont {B.~I.}\ \bibnamefont
  {Halperin}}, \bibinfo {author} {\bibfnamefont {Z.}~\bibnamefont {Te\ifmmode
  \check{s}\else \v{s}\fi{}anovi\ifmmode~\acute{c}\else \'{c}\fi{}}},\ and\
  \bibinfo {author} {\bibfnamefont {F.}~\bibnamefont {Axel}},\ }\bibfield
  {title} {\bibinfo {title} {Compatibility of crystalline order and the
  quantized hall effect},\ }\href {https://doi.org/10.1103/PhysRevLett.57.922}
  {\bibfield  {journal} {\bibinfo  {journal} {Phys. Rev. Lett.}\ }\textbf
  {\bibinfo {volume} {57}},\ \bibinfo {pages} {922} (\bibinfo {year}
  {1986})}\BibitemShut {NoStop}%
\bibitem [{\citenamefont {Kivelson}\ \emph {et~al.}(1986)\citenamefont
  {Kivelson}, \citenamefont {Kallin}, \citenamefont {Arovas},\ and\
  \citenamefont {Schrieffer}}]{kivelson1986}%
  \BibitemOpen
  \bibfield  {author} {\bibinfo {author} {\bibfnamefont {S.}~\bibnamefont
  {Kivelson}}, \bibinfo {author} {\bibfnamefont {C.}~\bibnamefont {Kallin}},
  \bibinfo {author} {\bibfnamefont {D.~P.}\ \bibnamefont {Arovas}},\ and\
  \bibinfo {author} {\bibfnamefont {J.~R.}\ \bibnamefont {Schrieffer}},\
  }\bibfield  {title} {\bibinfo {title} {Cooperative ring exchange theory of
  the fractional quantized hall effect},\ }\href
  {https://doi.org/10.1103/PhysRevLett.56.873} {\bibfield  {journal} {\bibinfo
  {journal} {Phys. Rev. Lett.}\ }\textbf {\bibinfo {volume} {56}},\ \bibinfo
  {pages} {873} (\bibinfo {year} {1986})}\BibitemShut {NoStop}%
\bibitem [{\citenamefont {Kivelson}\ \emph {et~al.}(1987)\citenamefont
  {Kivelson}, \citenamefont {Kallin}, \citenamefont {Arovas},\ and\
  \citenamefont {Schrieffer}}]{kivelson1987}%
  \BibitemOpen
  \bibfield  {author} {\bibinfo {author} {\bibfnamefont {S.}~\bibnamefont
  {Kivelson}}, \bibinfo {author} {\bibfnamefont {C.}~\bibnamefont {Kallin}},
  \bibinfo {author} {\bibfnamefont {D.~P.}\ \bibnamefont {Arovas}},\ and\
  \bibinfo {author} {\bibfnamefont {J.~R.}\ \bibnamefont {Schrieffer}},\
  }\bibfield  {title} {\bibinfo {title} {Cooperative ring exchange and the
  fractional quantum hall effect},\ }\href
  {https://doi.org/10.1103/PhysRevB.36.1620} {\bibfield  {journal} {\bibinfo
  {journal} {Phys. Rev. B}\ }\textbf {\bibinfo {volume} {36}},\ \bibinfo
  {pages} {1620} (\bibinfo {year} {1987})}\BibitemShut {NoStop}%
\bibitem [{\citenamefont {Te\ifmmode \check{s}\else
  \v{s}\fi{}anovi\ifmmode~\acute{c}\else \'{c}\fi{}}\ \emph
  {et~al.}(1989)\citenamefont {Te\ifmmode \check{s}\else
  \v{s}\fi{}anovi\ifmmode~\acute{c}\else \'{c}\fi{}}, \citenamefont {Axel},\
  and\ \citenamefont {Halperin}}]{halperin1989}%
  \BibitemOpen
  \bibfield  {author} {\bibinfo {author} {\bibfnamefont {Z.}~\bibnamefont
  {Te\ifmmode \check{s}\else \v{s}\fi{}anovi\ifmmode~\acute{c}\else
  \'{c}\fi{}}}, \bibinfo {author} {\bibfnamefont {F.~m.~c.}\ \bibnamefont
  {Axel}},\ and\ \bibinfo {author} {\bibfnamefont {B.~I.}\ \bibnamefont
  {Halperin}},\ }\bibfield  {title} {\bibinfo {title} {``hall crystal'' versus
  wigner crystal},\ }\href {https://doi.org/10.1103/PhysRevB.39.8525}
  {\bibfield  {journal} {\bibinfo  {journal} {Phys. Rev. B}\ }\textbf {\bibinfo
  {volume} {39}},\ \bibinfo {pages} {8525} (\bibinfo {year}
  {1989})}\BibitemShut {NoStop}%
\bibitem [{\citenamefont {Murthy}(2000)}]{ganpathy2000}%
  \BibitemOpen
  \bibfield  {author} {\bibinfo {author} {\bibfnamefont {G.}~\bibnamefont
  {Murthy}},\ }\bibfield  {title} {\bibinfo {title} {Hall crystal states at
  $\mathit{\ensuremath{\nu}}\phantom{\rule{0ex}{0ex}}=\phantom{\rule{0ex}{0ex}}2$
  and moderate landau level mixing},\ }\href
  {https://doi.org/10.1103/PhysRevLett.85.1954} {\bibfield  {journal} {\bibinfo
   {journal} {Phys. Rev. Lett.}\ }\textbf {\bibinfo {volume} {85}},\ \bibinfo
  {pages} {1954} (\bibinfo {year} {2000})}\BibitemShut {NoStop}%
\bibitem [{\citenamefont {Wilhelm}\ \emph {et~al.}(2021)\citenamefont
  {Wilhelm}, \citenamefont {Lang},\ and\ \citenamefont
  {L\"auchli}}]{wilhelm2021interplay}%
  \BibitemOpen
  \bibfield  {author} {\bibinfo {author} {\bibfnamefont {P.}~\bibnamefont
  {Wilhelm}}, \bibinfo {author} {\bibfnamefont {T.~C.}\ \bibnamefont {Lang}},\
  and\ \bibinfo {author} {\bibfnamefont {A.~M.}\ \bibnamefont {L\"auchli}},\
  }\bibfield  {title} {\bibinfo {title} {Interplay of fractional chern
  insulator and charge density wave phases in twisted bilayer graphene},\
  }\href {https://doi.org/10.1103/PhysRevB.103.125406} {\bibfield  {journal}
  {\bibinfo  {journal} {Phys. Rev. B}\ }\textbf {\bibinfo {volume} {103}},\
  \bibinfo {pages} {125406} (\bibinfo {year} {2021})}\BibitemShut {NoStop}%
\bibitem [{\citenamefont {Wilhelm}\ \emph {et~al.}(2023)\citenamefont
  {Wilhelm}, \citenamefont {Lang}, \citenamefont {Scheurer},\ and\
  \citenamefont {Läuchli}}]{Wilhelm_2023}%
  \BibitemOpen
  \bibfield  {author} {\bibinfo {author} {\bibfnamefont {P.}~\bibnamefont
  {Wilhelm}}, \bibinfo {author} {\bibfnamefont {T.}~\bibnamefont {Lang}},
  \bibinfo {author} {\bibfnamefont {M.}~\bibnamefont {Scheurer}},\ and\
  \bibinfo {author} {\bibfnamefont {A.}~\bibnamefont {Läuchli}},\ }\bibfield
  {title} {\bibinfo {title} {Non-coplanar magnetism, topological density wave
  order and emergent symmetry at half-integer filling of moiré chern bands},\
  }\bibfield  {journal} {\bibinfo  {journal} {SciPost Physics}\ }\textbf
  {\bibinfo {volume} {14}},\ \href
  {https://doi.org/10.21468/scipostphys.14.3.040}
  {10.21468/scipostphys.14.3.040} (\bibinfo {year} {2023})\BibitemShut
  {NoStop}%
\bibitem [{\citenamefont {Xu}\ \emph {et~al.}(2024)\citenamefont {Xu},
  \citenamefont {Li}, \citenamefont {Xu}, \citenamefont {Bi},\ and\
  \citenamefont {Zhang}}]{xu2024maximally}%
  \BibitemOpen
  \bibfield  {author} {\bibinfo {author} {\bibfnamefont {C.}~\bibnamefont
  {Xu}}, \bibinfo {author} {\bibfnamefont {J.}~\bibnamefont {Li}}, \bibinfo
  {author} {\bibfnamefont {Y.}~\bibnamefont {Xu}}, \bibinfo {author}
  {\bibfnamefont {Z.}~\bibnamefont {Bi}},\ and\ \bibinfo {author}
  {\bibfnamefont {Y.}~\bibnamefont {Zhang}},\ }\bibfield  {title} {\bibinfo
  {title} {Maximally localized {Wannier} functions, interaction models, and
  fractional quantum anomalous {Hall} effect in twisted bilayer {MoTe2}},\
  }\href {https://www.pnas.org/doi/10.1073/pnas.2316749121} {\bibfield
  {journal} {\bibinfo  {journal} {Proceedings of the National Academy of
  Sciences}\ }\textbf {\bibinfo {volume} {121}},\ \bibinfo {pages}
  {e2316749121} (\bibinfo {year} {2024})}\BibitemShut {NoStop}%
\bibitem [{\citenamefont {Reddy}\ \emph {et~al.}(2023)\citenamefont {Reddy},
  \citenamefont {Alsallom}, \citenamefont {Zhang}, \citenamefont {Devakul},\
  and\ \citenamefont {Fu}}]{reddy2023fractional}%
  \BibitemOpen
  \bibfield  {author} {\bibinfo {author} {\bibfnamefont {A.~P.}\ \bibnamefont
  {Reddy}}, \bibinfo {author} {\bibfnamefont {F.}~\bibnamefont {Alsallom}},
  \bibinfo {author} {\bibfnamefont {Y.}~\bibnamefont {Zhang}}, \bibinfo
  {author} {\bibfnamefont {T.}~\bibnamefont {Devakul}},\ and\ \bibinfo {author}
  {\bibfnamefont {L.}~\bibnamefont {Fu}},\ }\bibfield  {title} {\bibinfo
  {title} {Fractional quantum anomalous hall states in twisted bilayer
  ${\mathrm{mote}}_{2}$ and ${\mathrm{wse}}_{2}$},\ }\href
  {https://doi.org/10.1103/PhysRevB.108.085117} {\bibfield  {journal} {\bibinfo
   {journal} {Phys. Rev. B}\ }\textbf {\bibinfo {volume} {108}},\ \bibinfo
  {pages} {085117} (\bibinfo {year} {2023})}\BibitemShut {NoStop}%
\bibitem [{\citenamefont {Wang}\ \emph {et~al.}(2024)\citenamefont {Wang},
  \citenamefont {Zhang}, \citenamefont {Liu}, \citenamefont {He}, \citenamefont
  {Xu}, \citenamefont {Ran}, \citenamefont {Cao},\ and\ \citenamefont
  {Xiao}}]{wang2023fractional}%
  \BibitemOpen
  \bibfield  {author} {\bibinfo {author} {\bibfnamefont {C.}~\bibnamefont
  {Wang}}, \bibinfo {author} {\bibfnamefont {X.-W.}\ \bibnamefont {Zhang}},
  \bibinfo {author} {\bibfnamefont {X.}~\bibnamefont {Liu}}, \bibinfo {author}
  {\bibfnamefont {Y.}~\bibnamefont {He}}, \bibinfo {author} {\bibfnamefont
  {X.}~\bibnamefont {Xu}}, \bibinfo {author} {\bibfnamefont {Y.}~\bibnamefont
  {Ran}}, \bibinfo {author} {\bibfnamefont {T.}~\bibnamefont {Cao}},\ and\
  \bibinfo {author} {\bibfnamefont {D.}~\bibnamefont {Xiao}},\ }\bibfield
  {title} {\bibinfo {title} {Fractional chern insulator in twisted bilayer mote
  $ \_2$},\ }\href {https://doi.org/10.1103/PhysRevLett.132.036501} {\bibfield
  {journal} {\bibinfo  {journal} {Physical Review Letters}\ }\textbf {\bibinfo
  {volume} {132}},\ \bibinfo {pages} {036501} (\bibinfo {year}
  {2024})}\BibitemShut {NoStop}%
\bibitem [{\citenamefont {Goldman}\ \emph {et~al.}(2023)\citenamefont
  {Goldman}, \citenamefont {Reddy}, \citenamefont {Paul},\ and\ \citenamefont
  {Fu}}]{goldman2023zero}%
  \BibitemOpen
  \bibfield  {author} {\bibinfo {author} {\bibfnamefont {H.}~\bibnamefont
  {Goldman}}, \bibinfo {author} {\bibfnamefont {A.~P.}\ \bibnamefont {Reddy}},
  \bibinfo {author} {\bibfnamefont {N.}~\bibnamefont {Paul}},\ and\ \bibinfo
  {author} {\bibfnamefont {L.}~\bibnamefont {Fu}},\ }\bibfield  {title}
  {\bibinfo {title} {Zero-{Field} {Composite} {Fermi} {Liquid} in {Twisted}
  {Semiconductor} {Bilayers}},\ }\href
  {https://doi.org/10.1103/PhysRevLett.131.136501} {\bibfield  {journal}
  {\bibinfo  {journal} {Physical Review Letters}\ }\textbf {\bibinfo {volume}
  {131}},\ \bibinfo {pages} {136501} (\bibinfo {year} {2023})}\BibitemShut
  {NoStop}%
\bibitem [{\citenamefont {Dong}\ \emph
  {et~al.}(2023{\natexlab{a}})\citenamefont {Dong}, \citenamefont {Wang},
  \citenamefont {Ledwith}, \citenamefont {Vishwanath},\ and\ \citenamefont
  {Parker}}]{dong2023composite}%
  \BibitemOpen
  \bibfield  {author} {\bibinfo {author} {\bibfnamefont {J.}~\bibnamefont
  {Dong}}, \bibinfo {author} {\bibfnamefont {J.}~\bibnamefont {Wang}}, \bibinfo
  {author} {\bibfnamefont {P.~J.}\ \bibnamefont {Ledwith}}, \bibinfo {author}
  {\bibfnamefont {A.}~\bibnamefont {Vishwanath}},\ and\ \bibinfo {author}
  {\bibfnamefont {D.~E.}\ \bibnamefont {Parker}},\ }\bibfield  {title}
  {\bibinfo {title} {Composite {Fermi} {Liquid} at {Zero} {Magnetic} {Field} in
  {Twisted} {MoTe}\$\_2\$},\ }\href
  {https://doi.org/10.1103/PhysRevLett.131.136502} {\bibfield  {journal}
  {\bibinfo  {journal} {Physical Review Letters}\ }\textbf {\bibinfo {volume}
  {131}},\ \bibinfo {pages} {136502} (\bibinfo {year}
  {2023}{\natexlab{a}})}\BibitemShut {NoStop}%
\bibitem [{\citenamefont {Yu}\ \emph {et~al.}(2024)\citenamefont {Yu},
  \citenamefont {Herzog-Arbeitman}, \citenamefont {Wang}, \citenamefont
  {Vafek}, \citenamefont {Bernevig},\ and\ \citenamefont {Regnault}}]{Yu2023}%
  \BibitemOpen
  \bibfield  {author} {\bibinfo {author} {\bibfnamefont {J.}~\bibnamefont
  {Yu}}, \bibinfo {author} {\bibfnamefont {J.}~\bibnamefont
  {Herzog-Arbeitman}}, \bibinfo {author} {\bibfnamefont {M.}~\bibnamefont
  {Wang}}, \bibinfo {author} {\bibfnamefont {O.}~\bibnamefont {Vafek}},
  \bibinfo {author} {\bibfnamefont {B.~A.}\ \bibnamefont {Bernevig}},\ and\
  \bibinfo {author} {\bibfnamefont {N.}~\bibnamefont {Regnault}},\ }\bibfield
  {title} {\bibinfo {title} {Fractional chern insulators vs. non-magnetic
  states in twisted bilayer mote2},\ }\href
  {https://doi.org/10.1103/PhysRevB.109.045147} {\bibfield  {journal} {\bibinfo
   {journal} {Physical Review B}\ }\textbf {\bibinfo {volume} {109}},\ \bibinfo
  {pages} {045147} (\bibinfo {year} {2024})}\BibitemShut {NoStop}%
\bibitem [{\citenamefont {Abouelkomsan}\ \emph {et~al.}(2024)\citenamefont
  {Abouelkomsan}, \citenamefont {Reddy}, \citenamefont {Fu},\ and\
  \citenamefont {Bergholtz}}]{abouelkomsan2024band}%
  \BibitemOpen
  \bibfield  {author} {\bibinfo {author} {\bibfnamefont {A.}~\bibnamefont
  {Abouelkomsan}}, \bibinfo {author} {\bibfnamefont {A.~P.}\ \bibnamefont
  {Reddy}}, \bibinfo {author} {\bibfnamefont {L.}~\bibnamefont {Fu}},\ and\
  \bibinfo {author} {\bibfnamefont {E.~J.}\ \bibnamefont {Bergholtz}},\
  }\bibfield  {title} {\bibinfo {title} {Band mixing in the quantum anomalous
  {Hall} regime of twisted semiconductor bilayers},\ }\href
  {https://doi.org/10.1103/PhysRevB.109.L121107} {\bibfield  {journal}
  {\bibinfo  {journal} {Phys. Rev. B}\ }\textbf {\bibinfo {volume} {109}},\
  \bibinfo {pages} {L121107} (\bibinfo {year} {2024})}\BibitemShut {NoStop}%
\bibitem [{\citenamefont {Reddy}\ and\ \citenamefont
  {Fu}(2023)}]{reddy2023toward}%
  \BibitemOpen
  \bibfield  {author} {\bibinfo {author} {\bibfnamefont {A.~P.}\ \bibnamefont
  {Reddy}}\ and\ \bibinfo {author} {\bibfnamefont {L.}~\bibnamefont {Fu}},\
  }\bibfield  {title} {\bibinfo {title} {Toward a global phase diagram of the
  fractional quantum anomalous hall effect},\ }\href
  {https://doi.org/10.1103/PhysRevB.108.245159} {\bibfield  {journal} {\bibinfo
   {journal} {Phys. Rev. B}\ }\textbf {\bibinfo {volume} {108}},\ \bibinfo
  {pages} {245159} (\bibinfo {year} {2023})}\BibitemShut {NoStop}%
\bibitem [{\citenamefont {Qiu}\ \emph {et~al.}(2023)\citenamefont {Qiu},
  \citenamefont {Li}, \citenamefont {Luo},\ and\ \citenamefont
  {Wu}}]{qiu2023interaction}%
  \BibitemOpen
  \bibfield  {author} {\bibinfo {author} {\bibfnamefont {W.-X.}\ \bibnamefont
  {Qiu}}, \bibinfo {author} {\bibfnamefont {B.}~\bibnamefont {Li}}, \bibinfo
  {author} {\bibfnamefont {X.-J.}\ \bibnamefont {Luo}},\ and\ \bibinfo {author}
  {\bibfnamefont {F.}~\bibnamefont {Wu}},\ }\bibfield  {title} {\bibinfo
  {title} {Interaction-driven topological phase diagram of twisted bilayer
  ${\mathrm{mote}}_{2}$},\ }\href {https://doi.org/10.1103/PhysRevX.13.041026}
  {\bibfield  {journal} {\bibinfo  {journal} {Phys. Rev. X}\ }\textbf {\bibinfo
  {volume} {13}},\ \bibinfo {pages} {041026} (\bibinfo {year}
  {2023})}\BibitemShut {NoStop}%
\bibitem [{\citenamefont {Morales-Dur\'an}\ \emph {et~al.}(2024)\citenamefont
  {Morales-Dur\'an}, \citenamefont {Wei}, \citenamefont {Shi},\ and\
  \citenamefont {MacDonald}}]{morales2023magic}%
  \BibitemOpen
  \bibfield  {author} {\bibinfo {author} {\bibfnamefont {N.}~\bibnamefont
  {Morales-Dur\'an}}, \bibinfo {author} {\bibfnamefont {N.}~\bibnamefont
  {Wei}}, \bibinfo {author} {\bibfnamefont {J.}~\bibnamefont {Shi}},\ and\
  \bibinfo {author} {\bibfnamefont {A.~H.}\ \bibnamefont {MacDonald}},\
  }\bibfield  {title} {\bibinfo {title} {Magic angles and fractional chern
  insulators in twisted homobilayer transition metal dichalcogenides},\ }\href
  {https://doi.org/10.1103/PhysRevLett.132.096602} {\bibfield  {journal}
  {\bibinfo  {journal} {Phys. Rev. Lett.}\ }\textbf {\bibinfo {volume} {132}},\
  \bibinfo {pages} {096602} (\bibinfo {year} {2024})}\BibitemShut {NoStop}%
\bibitem [{\citenamefont {Sheng}\ \emph {et~al.}(2003)\citenamefont {Sheng},
  \citenamefont {Wan}, \citenamefont {Rezayi}, \citenamefont {Yang},
  \citenamefont {Bhatt},\ and\ \citenamefont {Haldane}}]{sheng2003chern}%
  \BibitemOpen
  \bibfield  {author} {\bibinfo {author} {\bibfnamefont {D.~N.}\ \bibnamefont
  {Sheng}}, \bibinfo {author} {\bibfnamefont {X.}~\bibnamefont {Wan}}, \bibinfo
  {author} {\bibfnamefont {E.~H.}\ \bibnamefont {Rezayi}}, \bibinfo {author}
  {\bibfnamefont {K.}~\bibnamefont {Yang}}, \bibinfo {author} {\bibfnamefont
  {R.~N.}\ \bibnamefont {Bhatt}},\ and\ \bibinfo {author} {\bibfnamefont
  {F.~D.~M.}\ \bibnamefont {Haldane}},\ }\bibfield  {title} {\bibinfo {title}
  {Disorder-driven collapse of the mobility gap and transition to an insulator
  in the fractional quantum hall effect},\ }\href
  {https://doi.org/10.1103/PhysRevLett.90.256802} {\bibfield  {journal}
  {\bibinfo  {journal} {Phys. Rev. Lett.}\ }\textbf {\bibinfo {volume} {90}},\
  \bibinfo {pages} {256802} (\bibinfo {year} {2003})}\BibitemShut {NoStop}%
\bibitem [{\citenamefont {Okamoto}\ \emph {et~al.}(2022)\citenamefont
  {Okamoto}, \citenamefont {Mohanta}, \citenamefont {Dagotto},\ and\
  \citenamefont {Sheng}}]{okamoto2022top}%
  \BibitemOpen
  \bibfield  {author} {\bibinfo {author} {\bibfnamefont {S.}~\bibnamefont
  {Okamoto}}, \bibinfo {author} {\bibfnamefont {N.}~\bibnamefont {Mohanta}},
  \bibinfo {author} {\bibfnamefont {E.}~\bibnamefont {Dagotto}},\ and\ \bibinfo
  {author} {\bibfnamefont {D.}~\bibnamefont {Sheng}},\ }\bibfield  {title}
  {\bibinfo {title} {Topological flat bands in a kagome lattice multiorbital
  system},\ }\href@noop {} {\bibfield  {journal} {\bibinfo  {journal}
  {Communications Physics}\ }\textbf {\bibinfo {volume} {5}},\ \bibinfo {pages}
  {198} (\bibinfo {year} {2022})}\BibitemShut {NoStop}%
\bibitem [{SM()}]{SM}%
  \BibitemOpen
  \href@noop {} {\bibinfo {title} {See supplementary materials for more
  supporting data.}}\BibitemShut {Stop}%
\bibitem [{\citenamefont {Kol}\ and\ \citenamefont
  {Read}(1993)}]{kol1993fractional}%
  \BibitemOpen
  \bibfield  {author} {\bibinfo {author} {\bibfnamefont {A.}~\bibnamefont
  {Kol}}\ and\ \bibinfo {author} {\bibfnamefont {N.}~\bibnamefont {Read}},\
  }\bibfield  {title} {\bibinfo {title} {Fractional quantum hall effect in a
  periodic potential},\ }\href {https://doi.org/10.1103/PhysRevB.48.8890}
  {\bibfield  {journal} {\bibinfo  {journal} {Phys. Rev. B}\ }\textbf {\bibinfo
  {volume} {48}},\ \bibinfo {pages} {8890} (\bibinfo {year}
  {1993})}\BibitemShut {NoStop}%
\bibitem [{\citenamefont {Polshyn}\ \emph {et~al.}(2021)\citenamefont
  {Polshyn}, \citenamefont {Zhang}, \citenamefont {Kumar}, \citenamefont
  {Soejima}, \citenamefont {Ledwith}, \citenamefont {Watanabe}, \citenamefont
  {Taniguchi}, \citenamefont {Vishwanath}, \citenamefont {Zaletel},\ and\
  \citenamefont {Young}}]{Polshyn_2021}%
  \BibitemOpen
  \bibfield  {author} {\bibinfo {author} {\bibfnamefont {H.}~\bibnamefont
  {Polshyn}}, \bibinfo {author} {\bibfnamefont {Y.}~\bibnamefont {Zhang}},
  \bibinfo {author} {\bibfnamefont {M.~A.}\ \bibnamefont {Kumar}}, \bibinfo
  {author} {\bibfnamefont {T.}~\bibnamefont {Soejima}}, \bibinfo {author}
  {\bibfnamefont {P.}~\bibnamefont {Ledwith}}, \bibinfo {author} {\bibfnamefont
  {K.}~\bibnamefont {Watanabe}}, \bibinfo {author} {\bibfnamefont
  {T.}~\bibnamefont {Taniguchi}}, \bibinfo {author} {\bibfnamefont
  {A.}~\bibnamefont {Vishwanath}}, \bibinfo {author} {\bibfnamefont {M.~P.}\
  \bibnamefont {Zaletel}},\ and\ \bibinfo {author} {\bibfnamefont {A.~F.}\
  \bibnamefont {Young}},\ }\bibfield  {title} {\bibinfo {title} {Topological
  charge density waves at half-integer filling of a moiré superlattice},\
  }\href {https://doi.org/10.1038/s41567-021-01418-6} {\bibfield  {journal}
  {\bibinfo  {journal} {Nature Physics}\ }\textbf {\bibinfo {volume} {18}},\
  \bibinfo {pages} {42–47} (\bibinfo {year} {2021})}\BibitemShut {NoStop}%
\bibitem [{\citenamefont {Lu}\ \emph {et~al.}(2023)\citenamefont {Lu},
  \citenamefont {Wu}, \citenamefont {Chen}, \citenamefont {Sun},\ and\
  \citenamefont {Meng}}]{lu2024}%
  \BibitemOpen
  \bibfield  {author} {\bibinfo {author} {\bibfnamefont {H.}~\bibnamefont
  {Lu}}, \bibinfo {author} {\bibfnamefont {H.-Q.}\ \bibnamefont {Wu}}, \bibinfo
  {author} {\bibfnamefont {B.-B.}\ \bibnamefont {Chen}}, \bibinfo {author}
  {\bibfnamefont {K.}~\bibnamefont {Sun}},\ and\ \bibinfo {author}
  {\bibfnamefont {Z.~Y.}\ \bibnamefont {Meng}},\ }\bibfield  {title} {\bibinfo
  {title} {From fractional quantum anomalous hall smectics to polar smectic
  metals: Nontrivial interplay between electronic liquid crystal order and
  topological order in correlated topological flat bands},\ }\href
  {https://arxiv.org/abs/2401.00363} {\bibfield  {journal} {\bibinfo  {journal}
  {arXiv preprint arXiv:2401.00363}\ } (\bibinfo {year} {2023})}\BibitemShut
  {NoStop}%
\bibitem [{\citenamefont {Kourtis}(2018)}]{kourtis2018symmetry}%
  \BibitemOpen
  \bibfield  {author} {\bibinfo {author} {\bibfnamefont {S.}~\bibnamefont
  {Kourtis}},\ }\bibfield  {title} {\bibinfo {title} {Symmetry breaking and the
  fermionic fractional chern insulator in topologically trivial bands},\ }\href
  {https://doi.org/10.1103/PhysRevB.97.085108} {\bibfield  {journal} {\bibinfo
  {journal} {Phys. Rev. B}\ }\textbf {\bibinfo {volume} {97}},\ \bibinfo
  {pages} {085108} (\bibinfo {year} {2018})}\BibitemShut {NoStop}%
\bibitem [{\citenamefont {Kourtis}\ and\ \citenamefont
  {Daghofer}(2014)}]{daghofer2014}%
  \BibitemOpen
  \bibfield  {author} {\bibinfo {author} {\bibfnamefont {S.}~\bibnamefont
  {Kourtis}}\ and\ \bibinfo {author} {\bibfnamefont {M.}~\bibnamefont
  {Daghofer}},\ }\bibfield  {title} {\bibinfo {title} {Combined topological and
  landau order from strong correlations in chern bands},\ }\href
  {https://doi.org/10.1103/PhysRevLett.113.216404} {\bibfield  {journal}
  {\bibinfo  {journal} {Phys. Rev. Lett.}\ }\textbf {\bibinfo {volume} {113}},\
  \bibinfo {pages} {216404} (\bibinfo {year} {2014})}\BibitemShut {NoStop}%
\bibitem [{\citenamefont {Dong}\ \emph
  {et~al.}(2023{\natexlab{b}})\citenamefont {Dong}, \citenamefont {Wang},
  \citenamefont {Wang}, \citenamefont {Soejima}, \citenamefont {Zaletel},
  \citenamefont {Vishwanath},\ and\ \citenamefont {Parker}}]{dong2023}%
  \BibitemOpen
  \bibfield  {author} {\bibinfo {author} {\bibfnamefont {J.}~\bibnamefont
  {Dong}}, \bibinfo {author} {\bibfnamefont {T.}~\bibnamefont {Wang}}, \bibinfo
  {author} {\bibfnamefont {T.}~\bibnamefont {Wang}}, \bibinfo {author}
  {\bibfnamefont {T.}~\bibnamefont {Soejima}}, \bibinfo {author} {\bibfnamefont
  {M.~P.}\ \bibnamefont {Zaletel}}, \bibinfo {author} {\bibfnamefont
  {A.}~\bibnamefont {Vishwanath}},\ and\ \bibinfo {author} {\bibfnamefont
  {D.~E.}\ \bibnamefont {Parker}},\ }\bibfield  {title} {\bibinfo {title}
  {Anomalous hall crystals in rhombohedral multilayer graphene i:
  Interaction-driven chern bands and fractional quantum hall states at zero
  magnetic field},\ }\href {https://arxiv.org/abs/2311.05568} {\bibfield
  {journal} {\bibinfo  {journal} {arXiv preprint arXiv:2311.05568}\ } (\bibinfo
  {year} {2023}{\natexlab{b}})}\BibitemShut {NoStop}%
\bibitem [{\citenamefont {Zhou}\ \emph {et~al.}(2023)\citenamefont {Zhou},
  \citenamefont {Yang},\ and\ \citenamefont {Zhang}}]{zhou2023fractional}%
  \BibitemOpen
  \bibfield  {author} {\bibinfo {author} {\bibfnamefont {B.}~\bibnamefont
  {Zhou}}, \bibinfo {author} {\bibfnamefont {H.}~\bibnamefont {Yang}},\ and\
  \bibinfo {author} {\bibfnamefont {Y.-H.}\ \bibnamefont {Zhang}},\ }\href@noop
  {} {\bibinfo {title} {Fractional quantum anomalous hall effects in
  rhombohedral multilayer graphene in the moir\'eless limit and in coulomb
  imprinted superlattice}} (\bibinfo {year} {2023}),\ \Eprint
  {https://arxiv.org/abs/2311.04217} {arXiv:2311.04217 [cond-mat.str-el]}
  \BibitemShut {NoStop}%
\end{thebibliography}%

\begin{widetext}
     		\renewcommand{\theequation}{S\arabic{equation}}
		\setcounter{equation}{0}
		\renewcommand{\thefigure}{S\arabic{figure}}
		\setcounter{figure}{0}
		\renewcommand{\thetable}{S\arabic{table}}
		\setcounter{table}{0}

\section{Supplemental Materials for: Quantum anomalous Hall crystal at fractional filling of moir\'e superlattices}

In the Supplemental Materials, we provide more numerical results to support the conclusions we have discussed in the main text. The details of the model and method for calculating Chern number are presented
in section A. Section B shows correlation functions  for smaller twist angle $\theta$ near the phase boundary of quantum anomalous Hall crystal (QAHC), and  Chern number results for larger systems at filling number $\nu=1/2$.   Section C has results for an alternative model by taking the
kinetic energy to zero to approach the ``flatband limit'' also at $\nu=1/2$.  Section D has results  to support a possible
QAHC phase for a different filling number $\nu=3/4$ at smaller twist angle $\theta\sim 1.6^\circ-1.8^\circ$.

\section{\label{sec:model} A. Model  and method}

 Focusing on twisted semiconductor bilayer $t$MoTe$_2$, 
 our continuum model following the Ref. \onlinecite{reddy2023fractional}  has the following form
\begin{equation}
    H_{\uparrow} = \begin{pmatrix}
    \frac{\hbar^2(-i \nabla - \kappa_{+})^2}{2m} + V_{+}(\mathbf{r})&& T(\mathbf{r}) \\ T^\dagger(\mathbf{r}) && \frac{\hbar^2(-i \nabla - \kappa_{-})^2}{2m} + V_{-}(\mathbf{r})
     \end{pmatrix}
\end{equation}

\begin{equation}
\begin{aligned}
V_{\pm} = -2V \sum_{i = 1,3,5} \cos(\mathbf{g}_i \cdot \mathbf{r}\pm \phi)\\
T(\mathbf{r}) = w ( 1 + e^{i \mathbf{g}_2 \cdot \mathbf{r}} +  e^{i \mathbf{g}_3 \cdot \mathbf{r}}).
\end{aligned}
\end{equation}
with $H_{\downarrow}$ its time reversal conjugate. $K$ points of the two layers are displaced due to the
interlayer twist and fold into the corners of the moir\'e Brillouin
zone, $\kappa_+$ and $\kappa_{-}$. $\mathbf{g}_i$ are 
 moir\'e reciprocal lattice
  vectors, $\mathbf{g}_i =\frac {4\pi} {{\sqrt 3} a_M } (cos \frac  {\pi (i-1)} 3,sin \frac {\pi (i-1)} 3)$. 
   $\kappa_+=({\bf g}_1+{\bf g}_2)/3$ and $\kappa_-=({\bf g}_1+{\bf g}_6)/3$.
  The moir\'e lattice constant is determined by twist angle $\theta$, 
  $a_M =\frac {a_0} {2sin(\theta/2)}$ with $a_0$ the atomic lattice constant.
We choose   $\phi=-91^\circ$, $V=11.2mev$,     $W=-13.3mev$,
$m=0.62m_e$, and $a_0=3.52$\cite{reddy2023fractional}  for   twisted  bilayer $t$MoTe$_2$
with tuning twist angle $\theta$. 

For most part of our simulations, we consider
spin polarized hole systems projected into the lowest energy band.
The long-range Coulomb interaction is considered with a dielectric constant
$\epsilon=10$. We diagonalize the many-body Hamiltonian in momentum space, and label energy eigenstates with momentum quantum number ${\bf k}=K_1{\bf T}_1+K_2{\bf T}_2$ with ${\bf T}_1, {\bf T}_2$  as unit vectors of crystal momentum.
We consider a finite size system with $N_s=N_1\times N_2$ unit cells and 
$N_1, N_2$ are the numbers of momentum points along ${\bf T}_1, {\bf T}_2$, respectively.

To probe the topological order of the many-body state, we calculate the many-body Chern number as an integral invariant of many-body wavefunction over twist boundary phase space (see Eq.3 of the main text).  
We discretize the boundary phase space into at least $12\times 12$ square meshes,
and calculate Berry phase $B_{ph}$ for   each square. 
This is done through calculating the  consecutive  wave function overlaps 
$B_{ph}(\phi_1, \phi_2)=arg(\prod_{i=1,4} \langle \psi_i|\psi_{i+1}\rangle)$ 
with  $|\psi_i>$ ($i$ is defined mod 4) representing four states at the four corners of the mesh square with the corresponding boundary phases $(\phi_1, \phi_2)$, $(\phi_1+\delta\phi,\phi_2)$, 
$(\phi_1+\delta\phi,\phi_2+\delta\phi)$, $(\phi_1, \phi_2+\delta\phi)$, respectively.
In calculating the overlap $\langle \psi_i|\psi_{i+1}\rangle$, the $\delta \phi$ should be small enough so that  $|B_{ph}(\phi_1,\phi_2)|<<\pi$ and
$|\langle \psi_i|\psi_{i+1}\rangle|\sim 1.0$.
We need to expand  $|\psi_i>$ into the original momentum basis for the continuum model, so that the contribution from the Berry curvature of the single particle orbitals in the flatband can be taken into account properly.
The total Chern number is obtained  as $C=\sum \frac {B_{ph}(\phi_1, \phi_2)} {2\pi}$   over all the mesh squares in the 
$2\pi \times 2\pi$ boundary phase space.

\begin{figure}
  \includegraphics[width=0.7\linewidth]{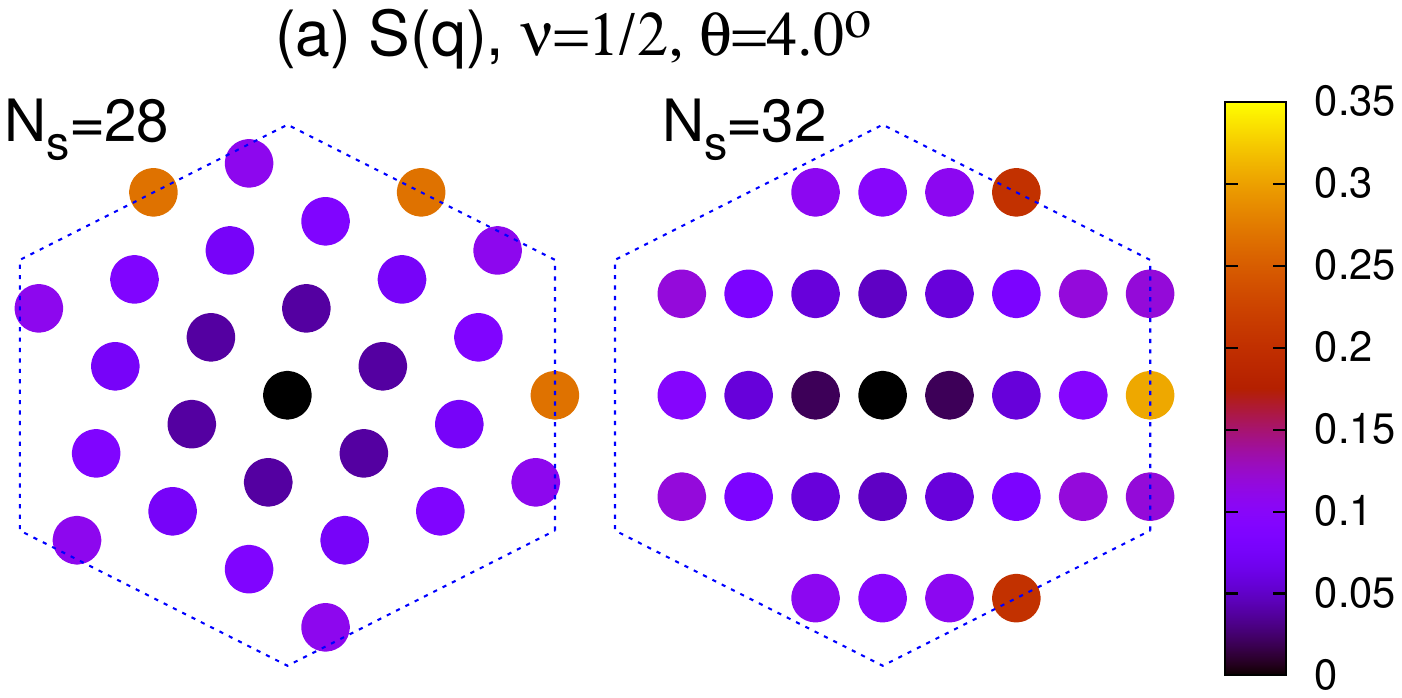}\\
  \vspace{-4.0cm}
  \includegraphics[width=0.7\linewidth]{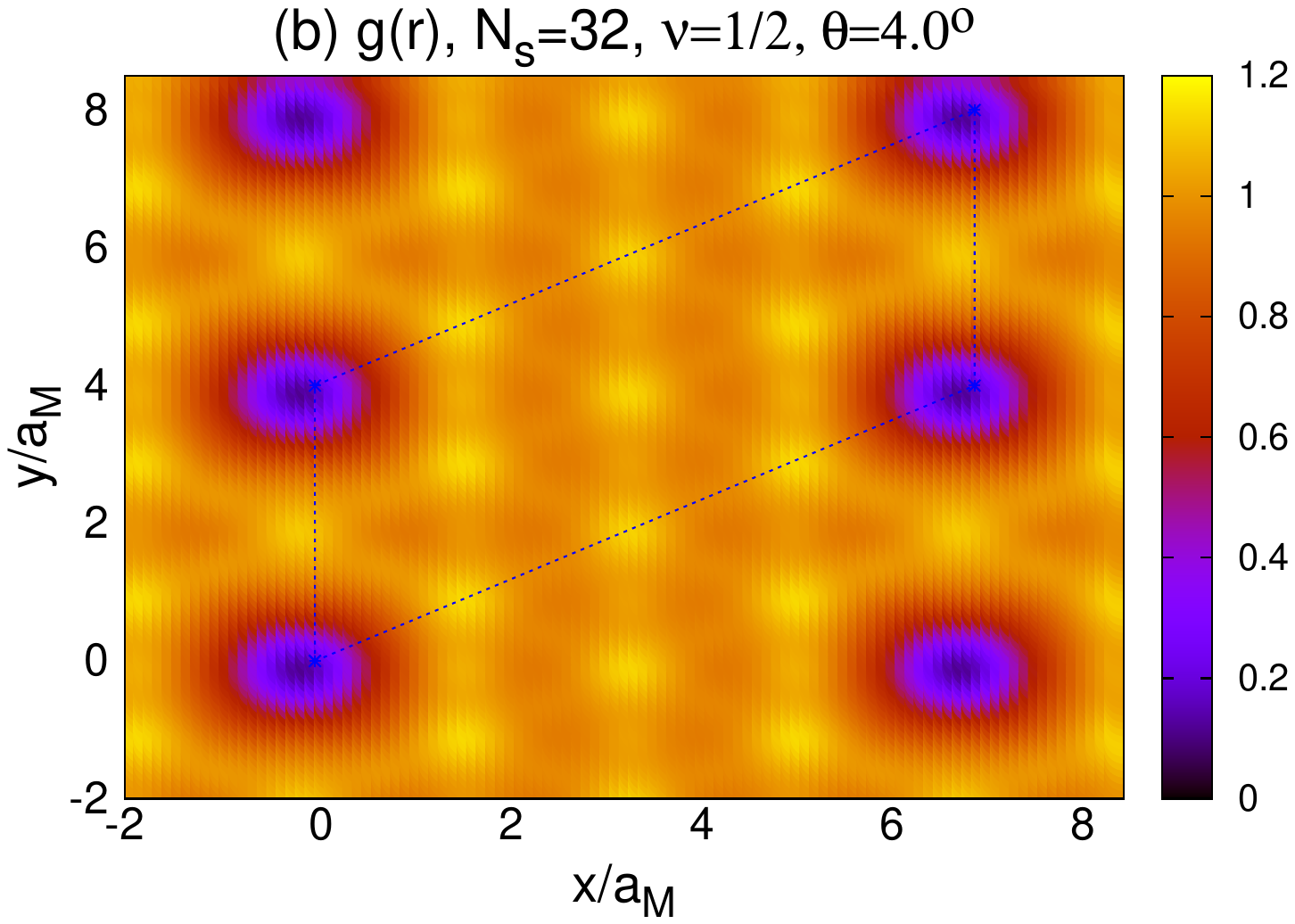}
    \vspace*{-1.0cm}
\caption{(a) Projected density structure factor $S(q)$ for
$N_s=28$ and 32; (b) Pair correlation function $g(r)$ 
for $N_s=32$ at $\nu=1/2$ and $\theta=4.0^{\circ}$ for the ground state at $\Gamma$ point.  All other near degenerating ground states show the same feature.
} 
\label{fig:figS1_sq}
\end{figure}

\begin{figure}
   \includegraphics[width=0.48\linewidth]{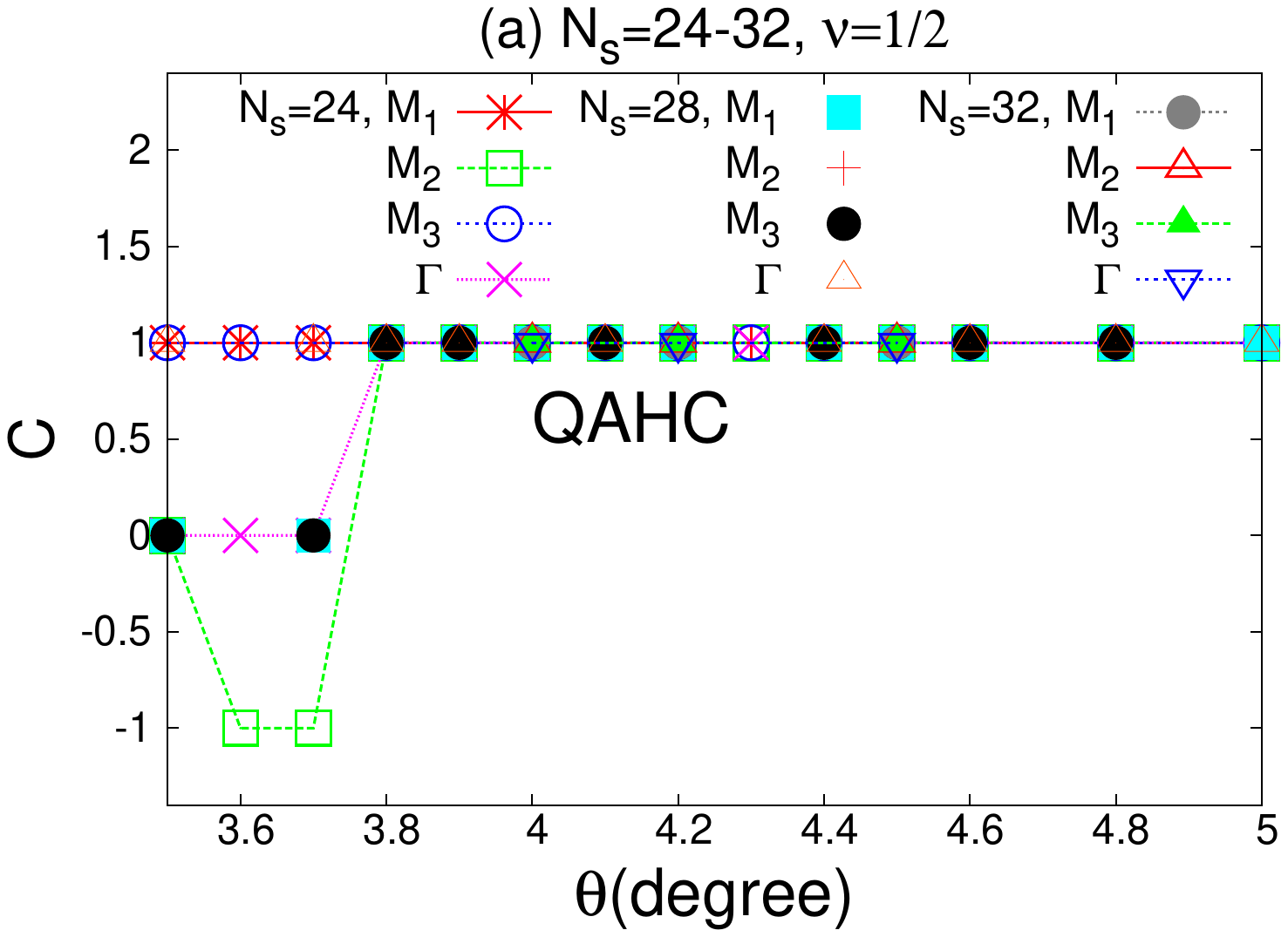}
    \includegraphics[width=0.48\linewidth]{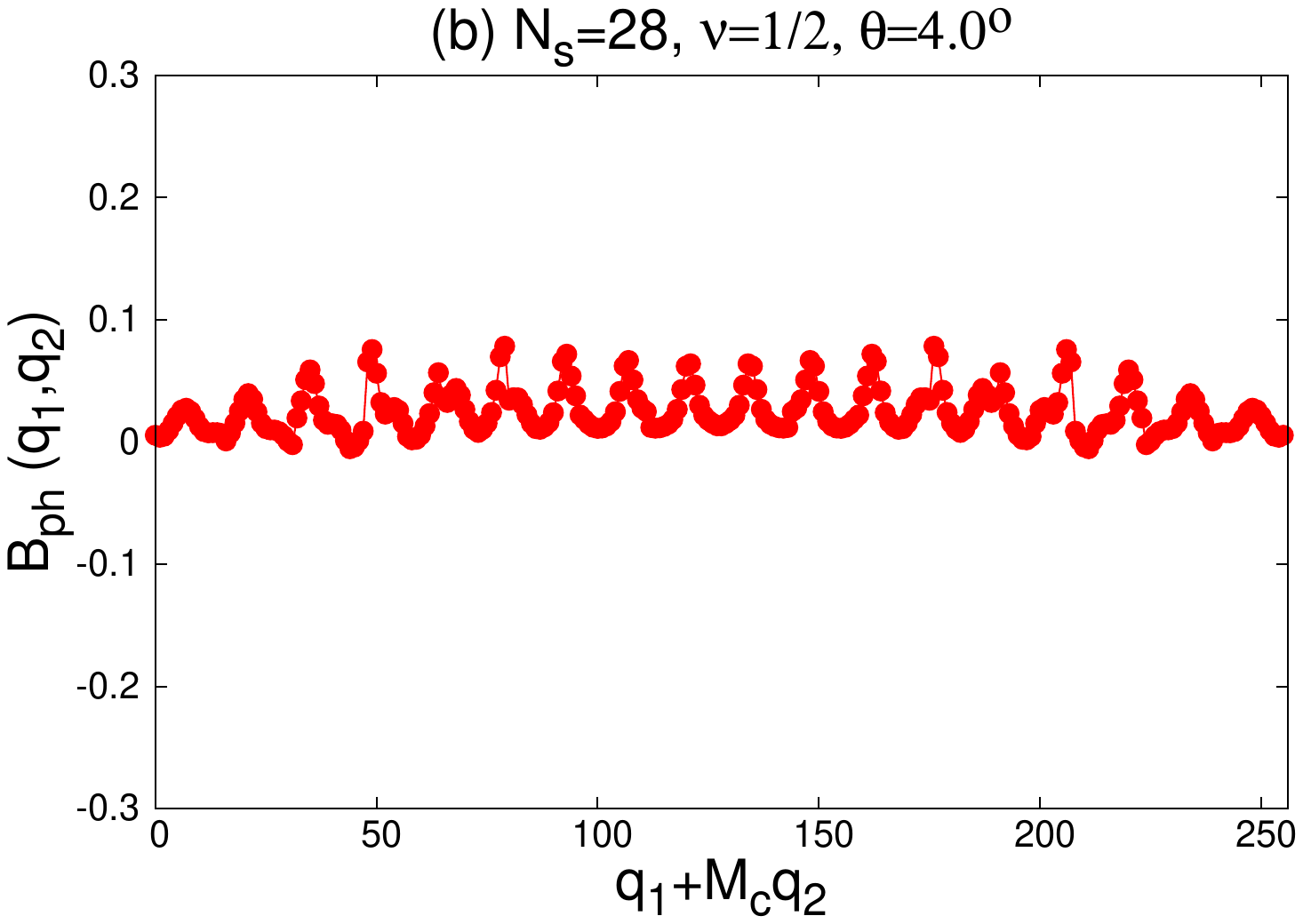} \\
    \vspace{-0.80cm}
    \includegraphics[width=0.48\linewidth]{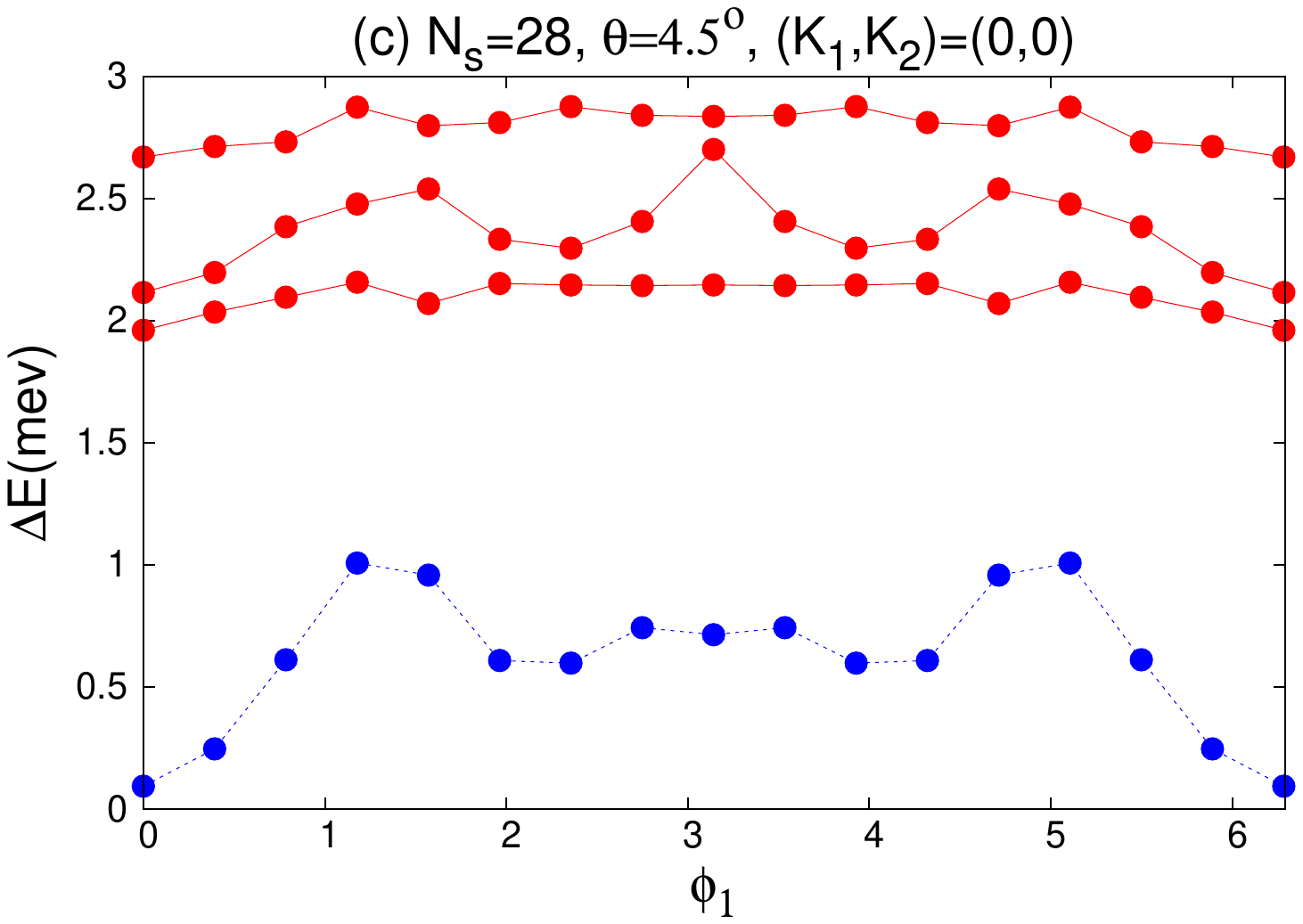}
    \includegraphics[width=0.48\linewidth]{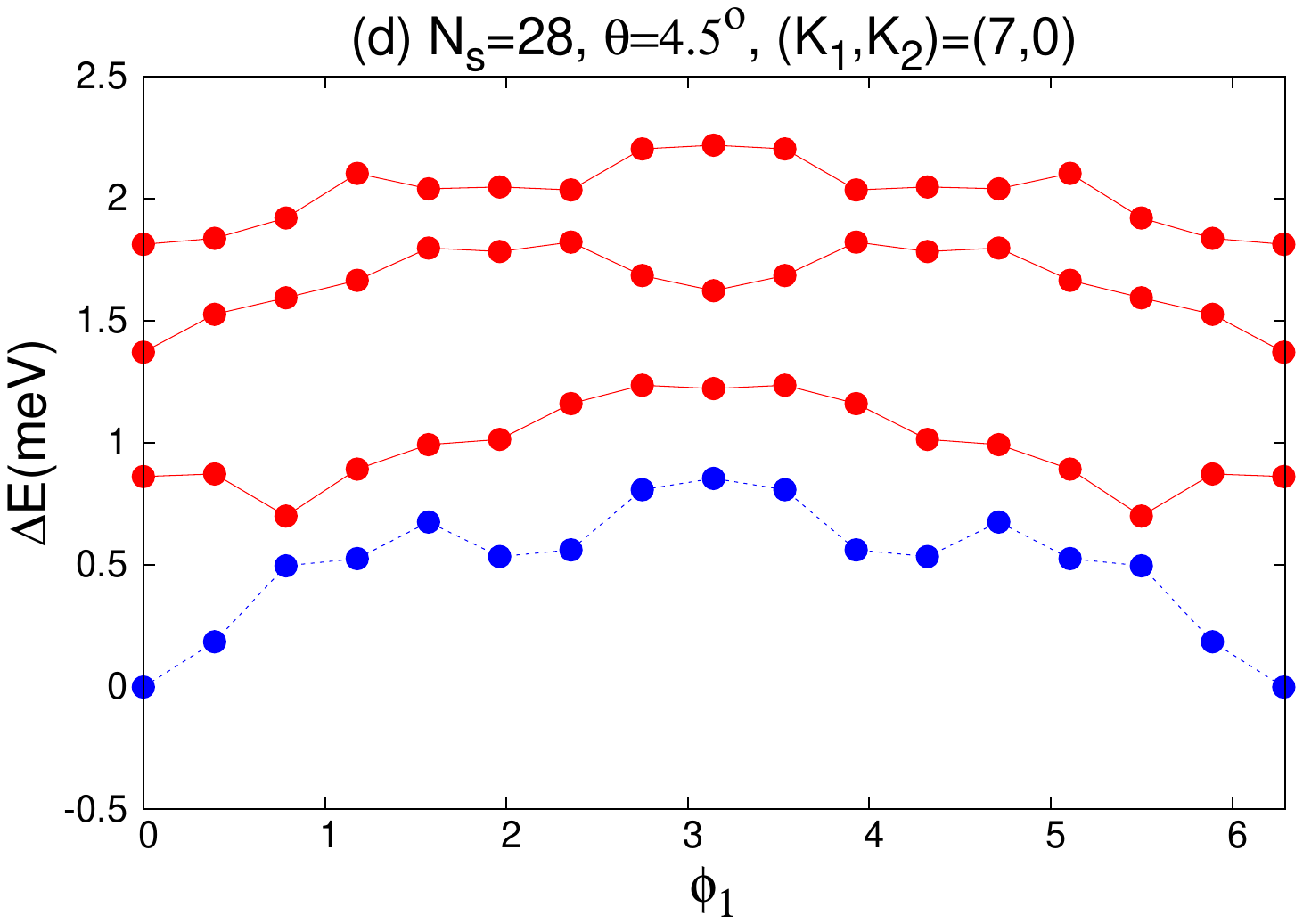} 
    \vspace*{-0.50cm}
\caption{(a) Chern numbers of  lowest energy states with different momenta versus $\theta$ for $N_s=24$, 28 and 32 systems at filling $\nu=1/2$ in the QAHC phase.
The QAHC state starts to emerge at $\theta_c\sim 3.8$ as we observe $C=1$
for all the ground states in these systems.
(b) Berry phase from each square in boundary phase space with $(\phi_1,\phi_2)=(q_1/M_c, q_2/M_c)$ in units of $2\pi$ for $\theta=4.0^{\circ}$. Here $M_c=16$ and $M_c\times M_c$ is the number of mesh squares for Chern number calculations.
(c-d) Energy levels versus the inserted flux $\phi_1$ for two momentum sectors $(K_1,K_2)=(0,0)$ (c)
and $(K_1,K_2)=(7,0)$ (d), respectively.} 
\label{fig:figS2_chern}
\end{figure}

\section{\label{sec:nu12}B.  Additional results for $\nu=1/2$ QAHC}

In this section, we present  results for the QAHC phase for density correlations
around phase boundary at $\theta=4.0^\circ$ and more results for the ground state Chern number 
with different cluster sizes $N_s=24-32$.  
As shown in Fig. \ref{fig:figS1_sq}(a), the projected density structure factor $S({\bf q})$
show prominent peaks around three $M$ points, which grows slightly with the increase of system size
from $N_s=28$ to 32.  The pair correlations (Fig. \ref{fig:figS1_sq}(b))
 demonstrate $2\times 2 $ crystal order at $\theta=4.0^\circ$, suggesting the system already has a transition into a crystal phase.

 To identify topological order, we have focused on two lowest states averaged Chern number in the main text, which allows us to follow the evolution of composite Fermi liquid (CFL) into the transition regime.
 However,  because our crystal phase has a gap separating the near degenerating ground states from excited states, thus the topological order at Hall crystal side can be established from  the Chern numbers of the near degenerating ground states.
 In Fig. \ref{fig:figS2_chern}(a), we show the ground state Chern number in each of
 the ground state momentum sectors including $M_i$ ($i=1-3$) and $\Gamma$ points for the range of $\theta=3.5^\circ-5.0^\circ$. We see that the fluctuation Chern numbers
 appear between $\theta=3.6^\circ-3.7^\circ$, while at smaller $\theta=3.5^\circ$, the averaged
 Chern number goes to $0.5$ consistent with CFL.
 Staring from $\theta\geq 3.8^\circ$, the ground state has $C=1$ for all momentum sectors 
 and $N_s=24-32$, suggesting more accurate transition point
 $\theta_c\sim 3.8^\circ$ for entering the QAHC phase. 
 To obtain these Chern numbers accurately, we divide the boundary phase into
 $16\times 16$ square meshes, with  Berry phase $B_{ph}$ of each square shown in Fig. \ref {fig:figS2_chern}(b), which is a smooth function of boundary phase
 $(\phi_1, \phi_2)$ with very small magnitude fluctuations.   
 The many-body energy spectra show smooth variation  with boundary phases, and the adjacent energy gaps remain open as shown in Fig. \ref{fig:figS2_chern}(c-d) for states in different momentum sectors, which allow a well defined Chern number for these low energy states. 
 
 \begin{figure}
   \includegraphics[width=0.48\linewidth]{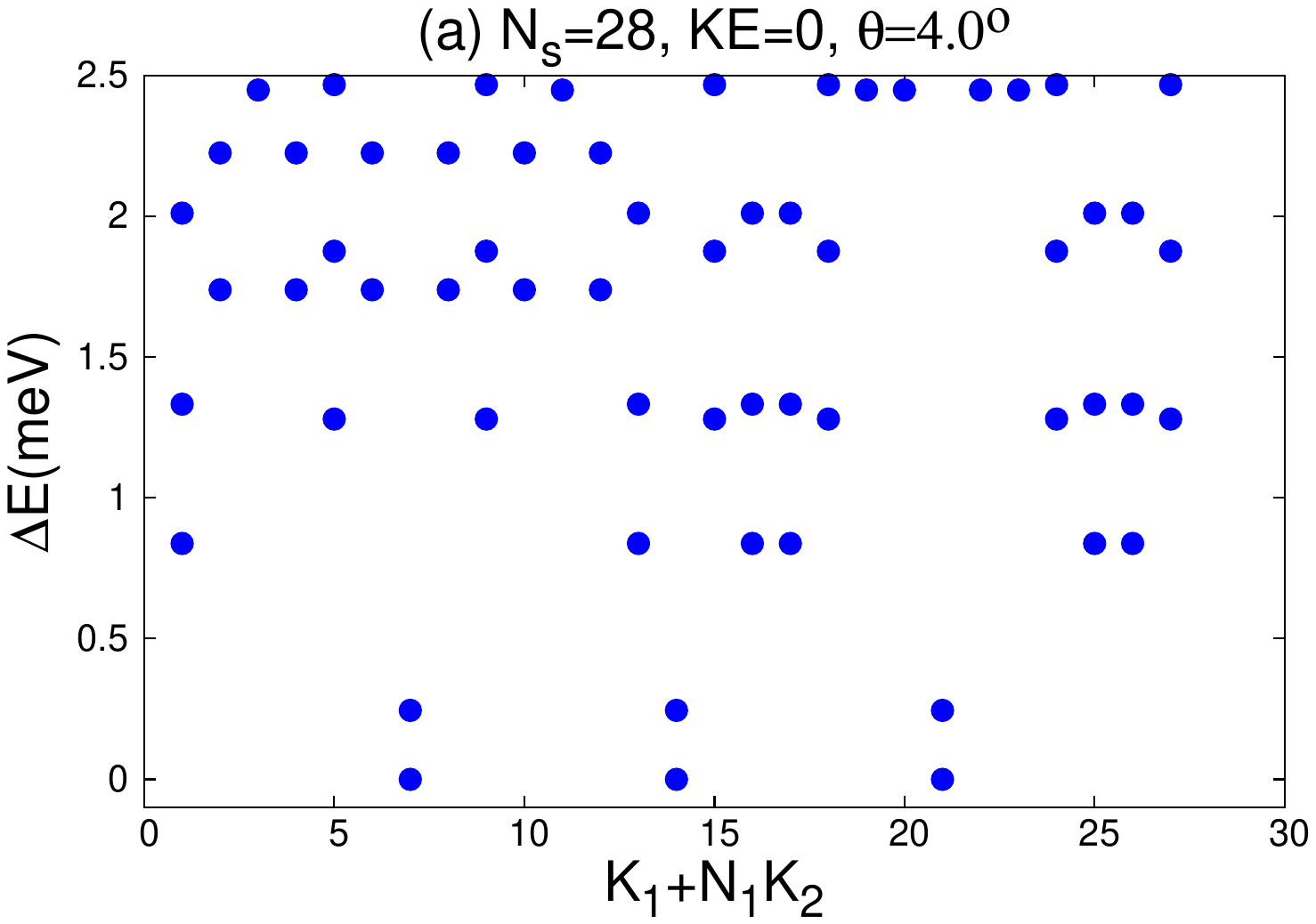}
    \includegraphics[width=0.48\linewidth]   {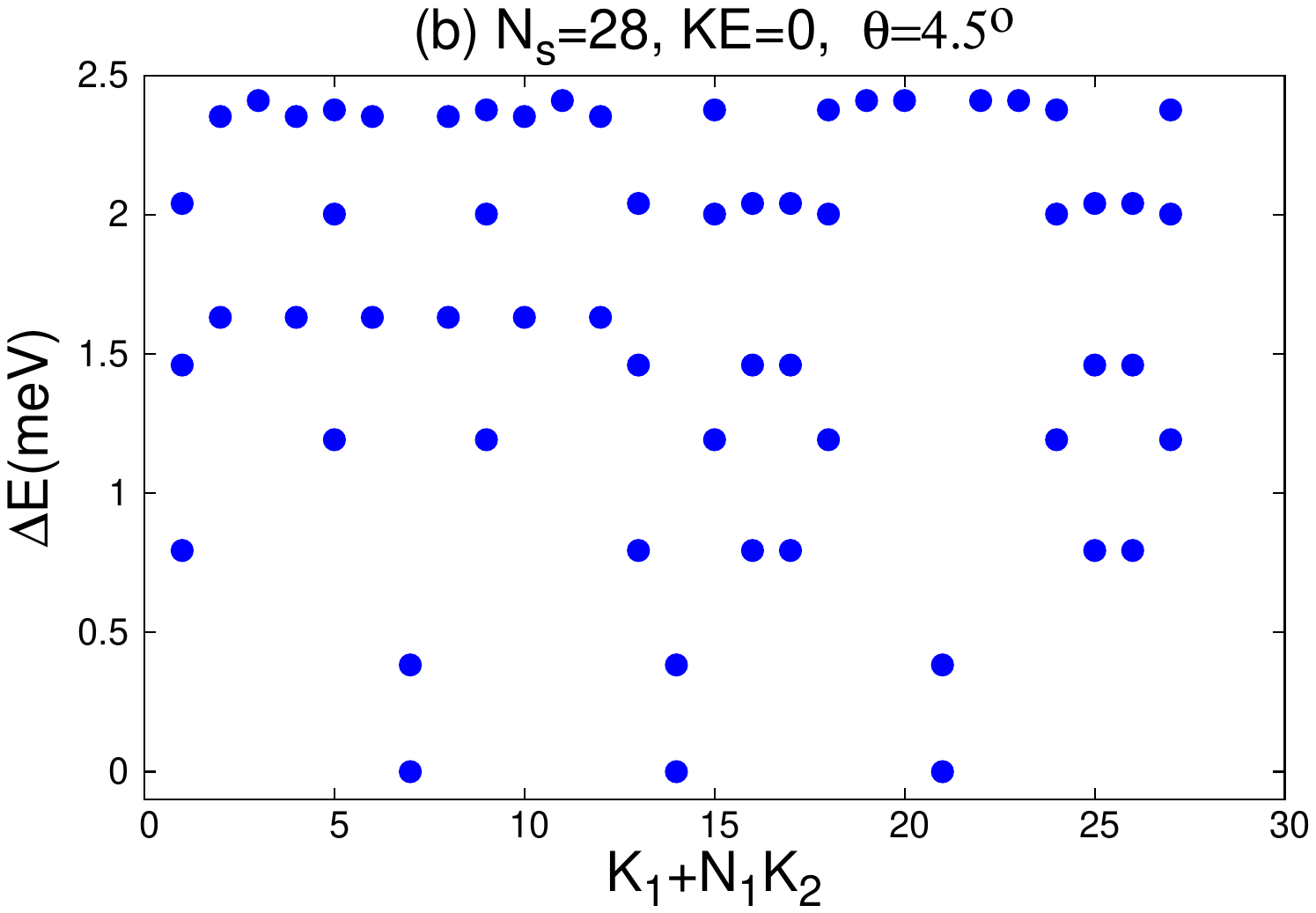}\\ 
   \vspace{-0.65cm}
    \caption{(a-b)Energy spectra for flatband limit ($KE=0$) by setting single particle energies to zero for all orbitals of the lowest energy band for holes
    at $N_s=28$, $\theta=4.0^{\circ}$, $4.5^{\circ}$ and filling number $\nu=1/2$.}
    \label{fig:figS3_eng_KE0}
\end{figure}

\begin{figure}    
    \includegraphics[width=0.48\linewidth]
   {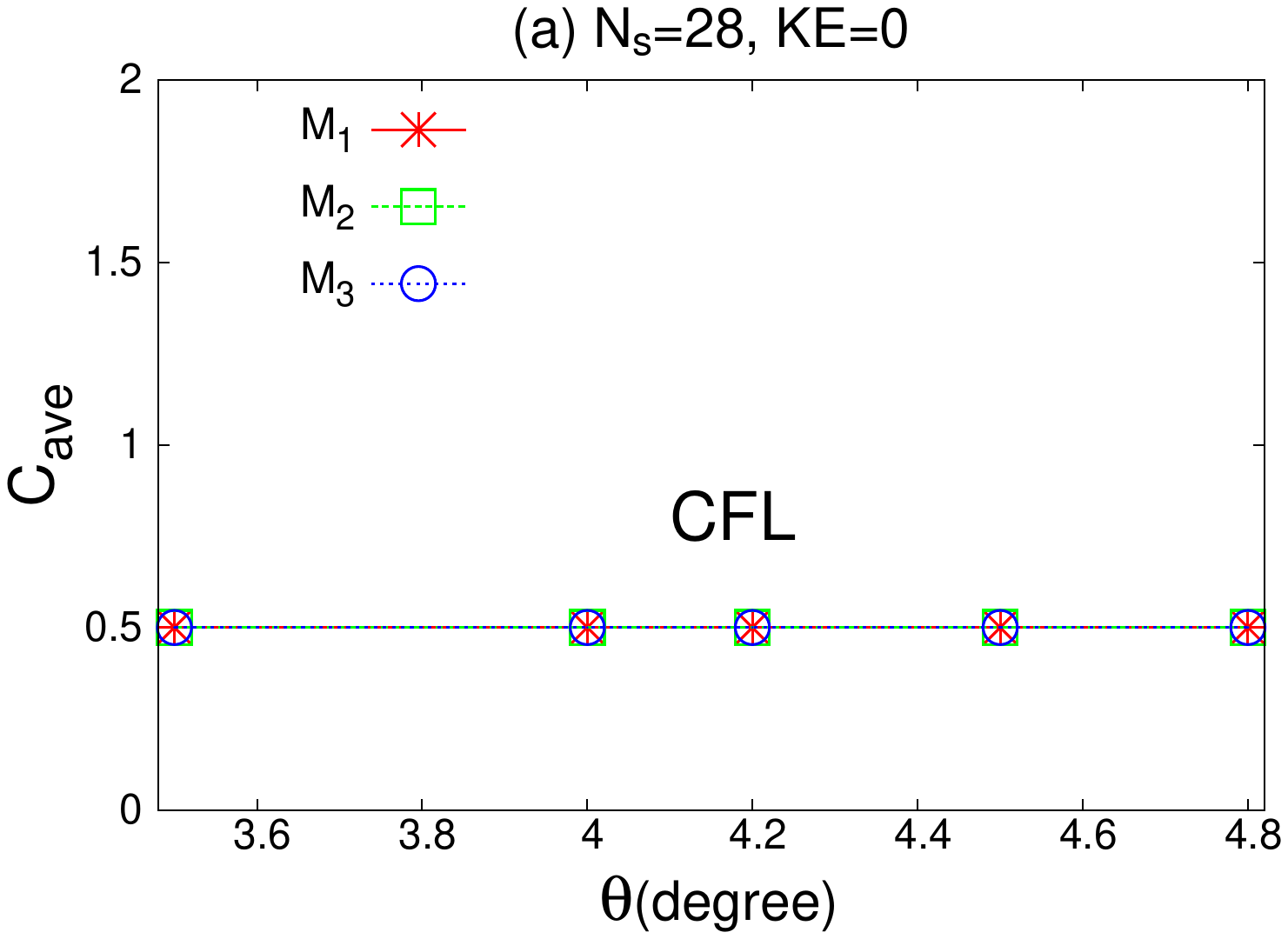}
   \includegraphics[width=0.48\linewidth]
   {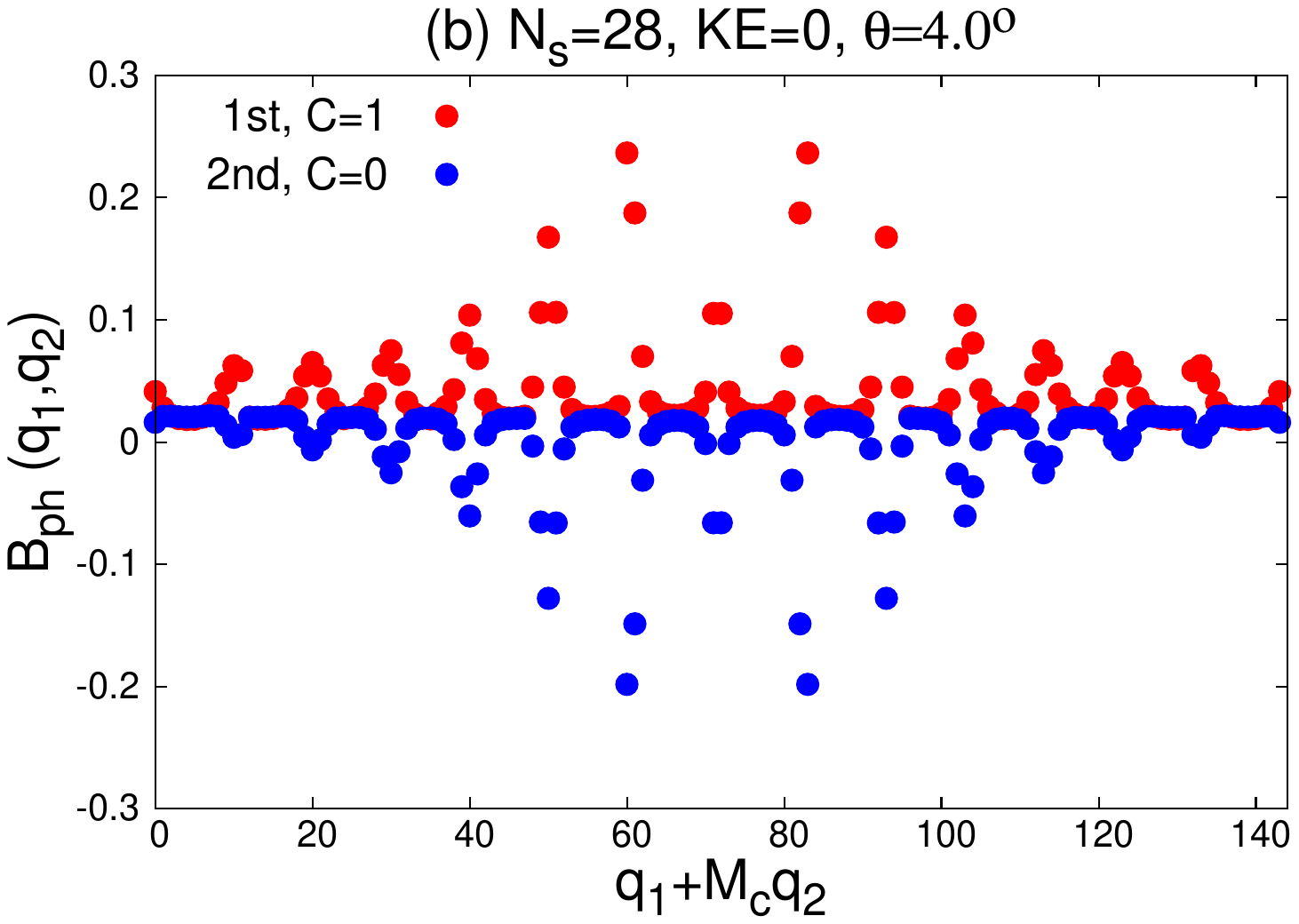}
   \vspace{-0.6cm}
    \caption{(a-b)The averaged Chern number of lowest two states in ground state momentum sectors at $\nu=1/2$. (d) Berry phase from each square in boundary phase space with $(\phi_1,\phi_2)=(q_1/M_c, q_2/M_c)$ in units of $2\pi$ for $\theta=4.0^{\circ}$ at $\nu=1/2$. Here $M_c=12$ and we use $12\times 12$  mesh squares for Chern number calculations. }
\label{fig:figS4_chern_KE0}
\end{figure}

\section{\label{sec:model_flat}C.  Quantum phase for the flatband limit for $\nu=1/2$ system}

To provide more intuitive  understanding of  the driving force for the emergence of the QAHC phase in twisted
moir\`e system, we study the same model at the flatband limit by switching off its kinetic
energy of each orbital in the lowest energy band.   The energy spectrum of the ideal
flatband is shown in Fig. \ref{fig:figS3_eng_KE0} for $\theta=4.0^{\circ}$ and $4.5^{\circ}$, respectively.   In both cases, there are a pair of lowest energy states in each of the three $M$ points,
with small  energy separation between them. The averaged Chern number (Fig. \ref{fig:figS4_chern_KE0}(a)) of these near degenerating
ground states $C_{ave}=1/2$ for two levels in the same momentum sector. These a pair of states also show correlated Berry curvature, with the fluctuating Berry phases canceling each other as illustrated in Fig. \ref{fig:figS4_chern_KE0}(b). 
These results are consistent with a CFL phase in the whole regime of $\theta=3.5^\circ-4.5^\circ$ in the absence of kinetic energy.   Thus, we find that the interplay
between the kinetic energy and correlation effect  is important for
the emergence of the QAHC in the original model shown in the main text.

\section{\label{sec:nu34}D. A possible quantum Anomalous  Hall crystal at filling number $\nu=3/4$}
In this section, we present our numerical study for the  filling number
$\nu=3/4$, where previous study found CFL\cite{reddy2023toward} phase near magic angle $\theta=2.0^{\circ}$. 
We first look at the energy spectra of the system at $\theta=1.8^{\circ}$ for 
$N_s=28$ and 32 as shown in Fig. \ref{fig:figS5_eng_nu34}.   For $C_6$ rotational invariant cluster $N_s=28$, we identify 
 4-fold ground state near degeneracy consistent with a CDW order.   For $N_s=32$, we identify two lowest energy states
 at $\Gamma$ and $M$ point consistent with a stripe order.

As shown in Fig.~\ref{fig:figS6_sq_nu34}, $S({\bm q})$ shows strong circular-like  peaks around three $M$ points and nearby points 
for $N_s=28$.   However, for anisotropic system $N_s=32$,
we find a much enhanced peak at one of the M point corresponding to an order momentum 
$\pi, 0$, consistent with a $2x1$ stripe phase.
These results indicate while crystal order is strong  for this system, their
order patterns are still tunable by the finite size effects of different clusters.

\begin{figure}
 \includegraphics[width=0.48\linewidth]{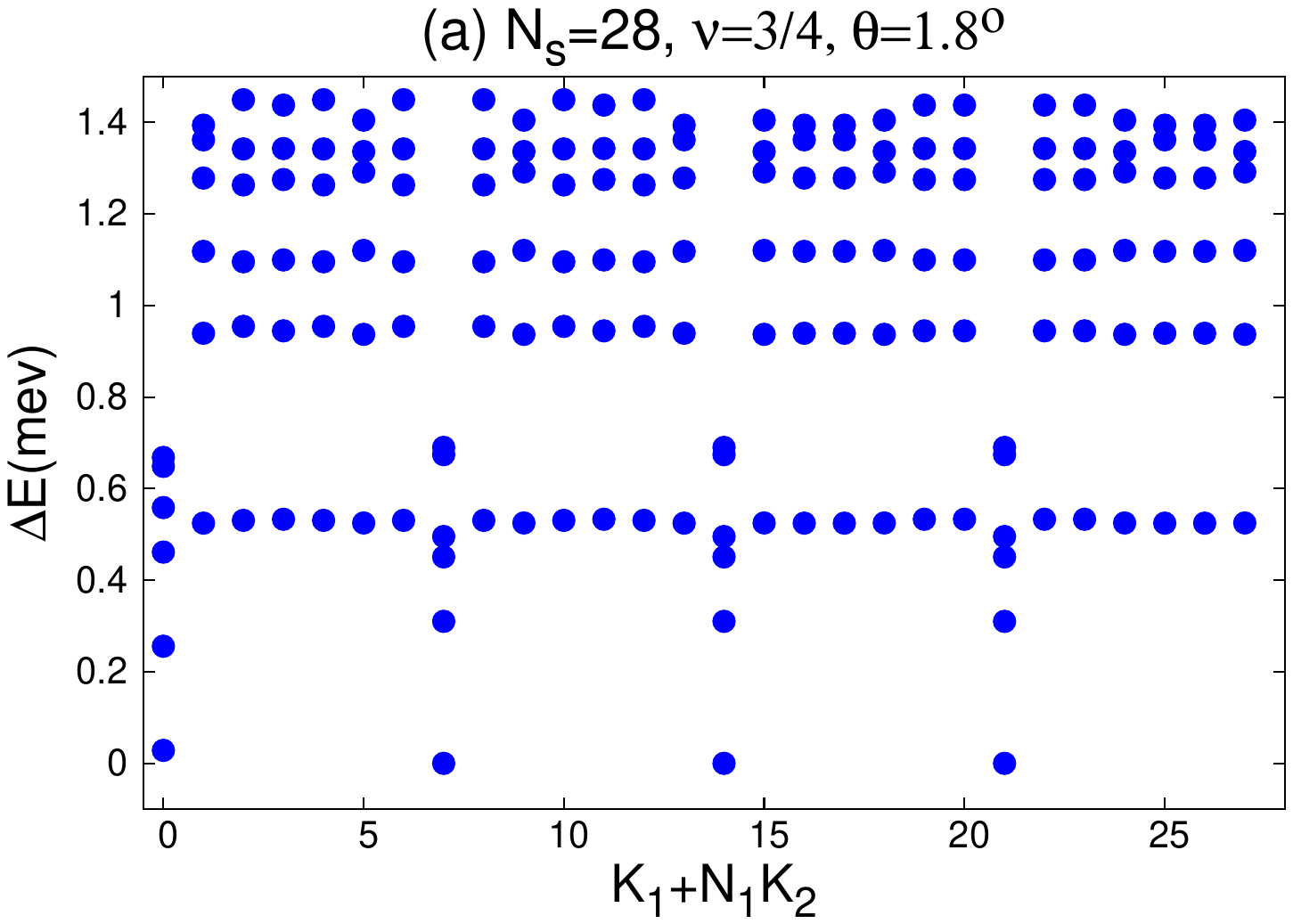} \includegraphics[width=0.48\linewidth]
 {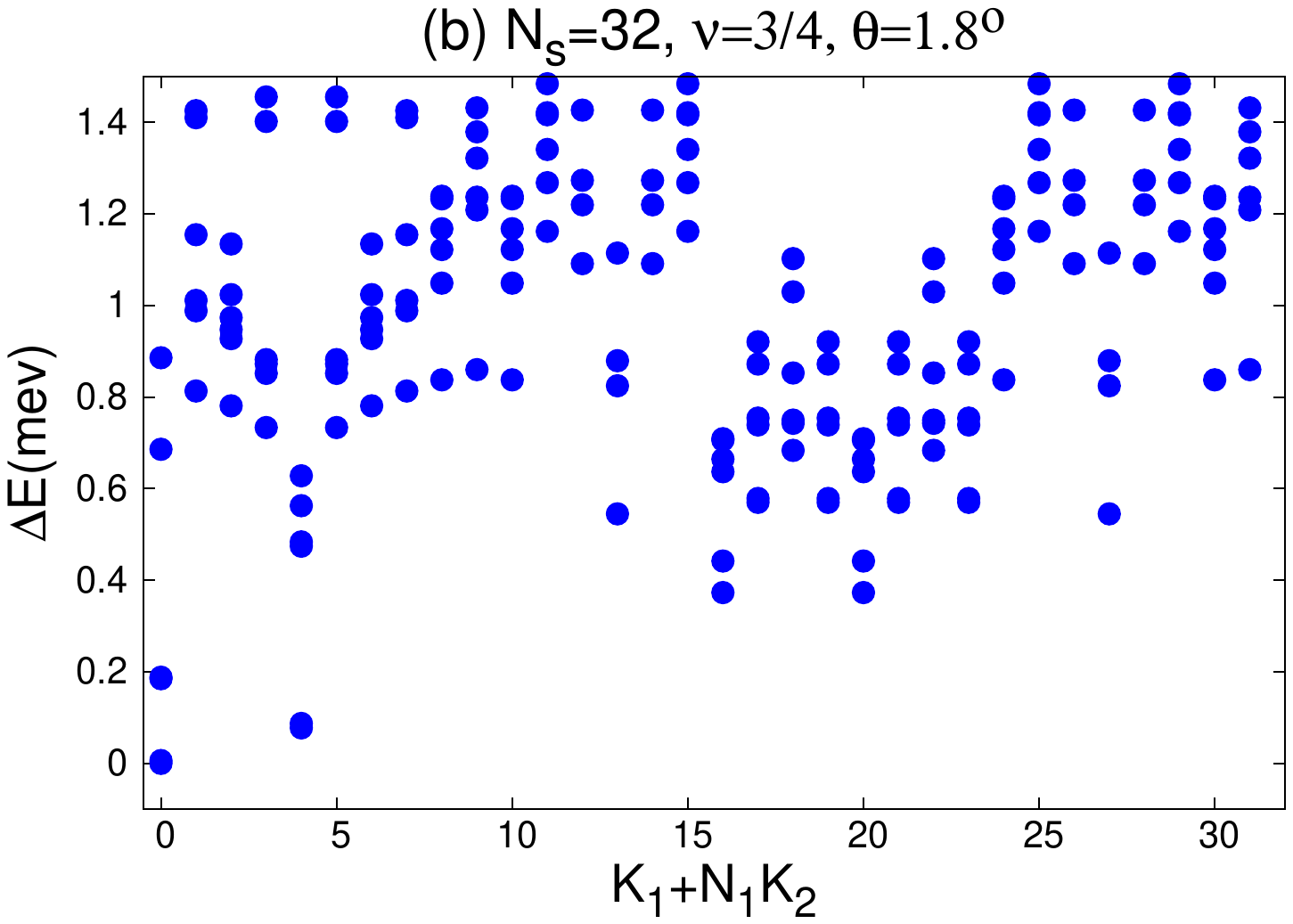}
 \caption{Energy spectra versus momentum for $\nu=3/4$ and $\theta=1.8^{\circ}$ 
 for system size $N_s=28$ and 32.  For $C_6$ rotational invariant cluster $N_s=28$, we identify 
 4-fold near degeneracy consistent with CDW order.   For $N_s=32$, we identify two lowest energy states
 consistent with a stripe order.} 
\label{fig:figS5_eng_nu34}
 \end{figure}

 \begin{figure} 
\includegraphics[width=0.8\linewidth,height=0.67\linewidth]{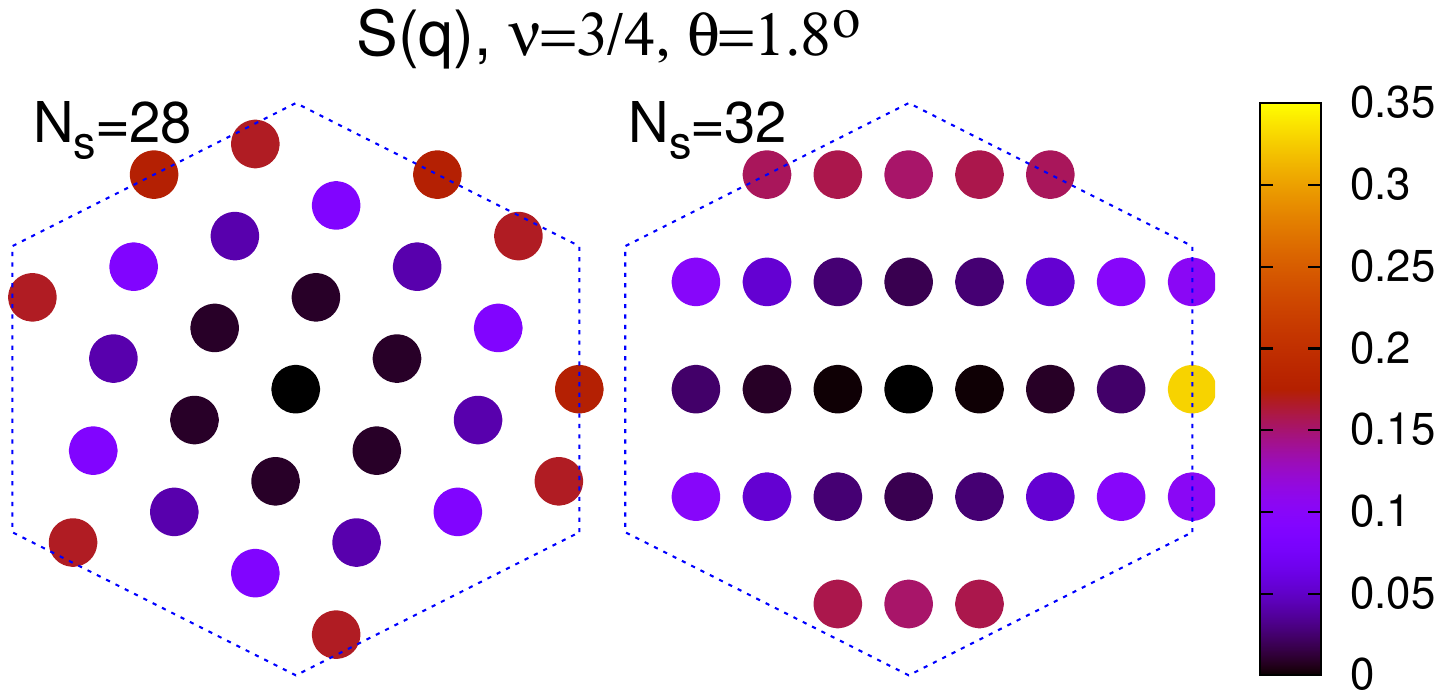}
\vspace{-4.200cm}
\caption{The projected density structure factor for $N_s=28$ and $32$
 at $\theta=1.8^{\circ}$.} 
\label{fig:figS6_sq_nu34}
\end{figure}

 \begin{figure}
 \includegraphics[width=0.8\linewidth]{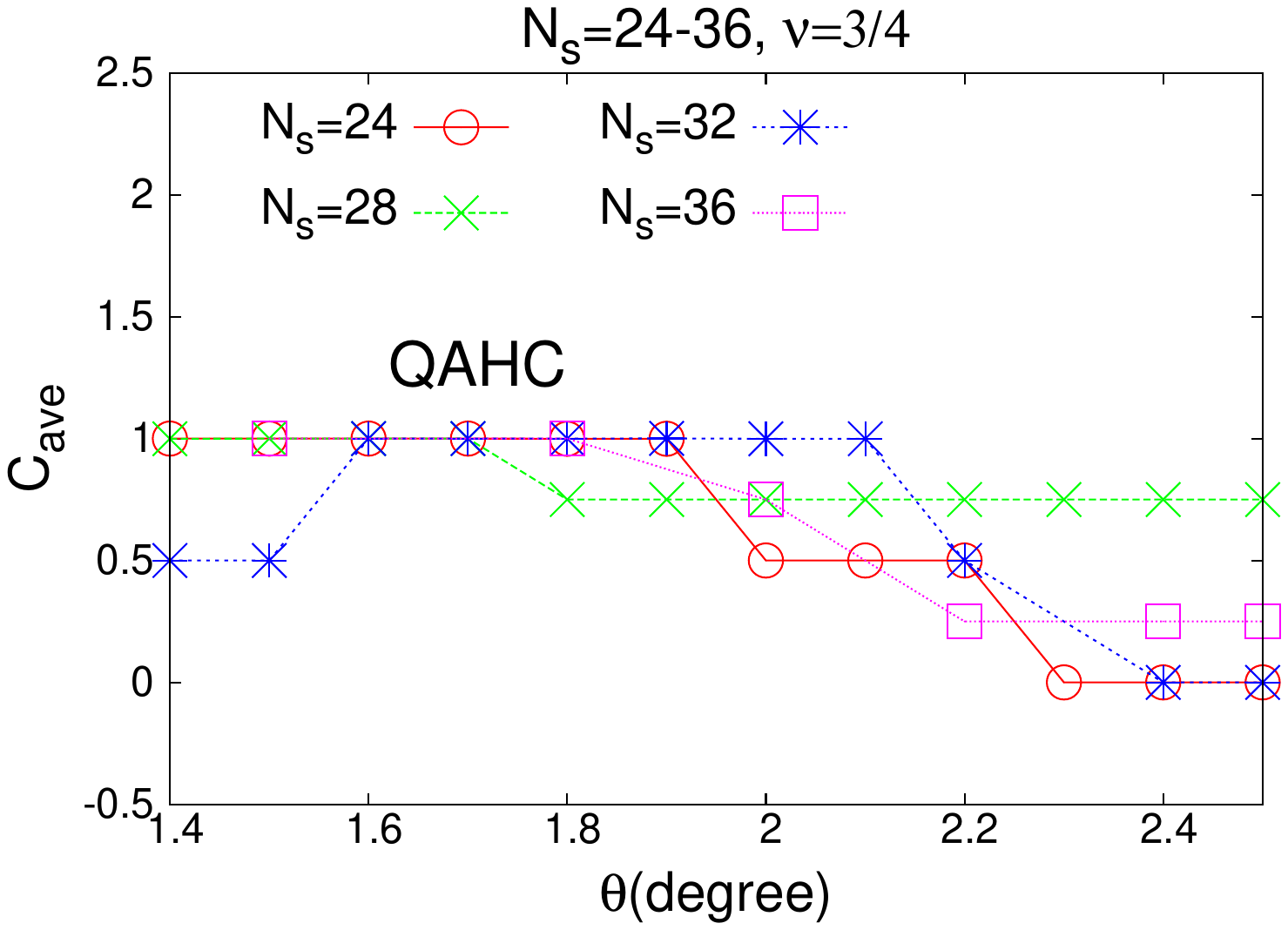}
    \vspace*{-1.0cm}
\caption{Chern numbers averaged over  lowest energy states with different momenta
when they are connected through inserting flux,  versus $\theta$ for $N_s=24-36$ systems at filling $\nu=3/4$. Quantized $C_{ave}=1$ is identified near $\theta=1.6^{\circ}-1.8^{\circ}$ for larger systems
$N_s=32$ and 36, consistent with a possible QAHC state.} 
\label{fig:figS7_chern_nu34}
\end{figure}

To identify topological order, we calculate the averaged Chern number for different ground state momentum sectors. The averaging is necessary for 
$\nu=3/4$ for clusters such as $N_s=28$ and $36$, as the ground state momentum
will be changed by $\Delta {\bf K}=(N_e {\bf T}_1, N_e {\bf T}_2)$ through inserting one flux quantum along both $L_1$ and $L_2$
directions.
 In Fig. \ref{fig:figS7_chern_nu34}, we show  $C_{ave}=1$ for 
 twist angle between $\theta=1.6^\circ-1.8^\circ$ for
 larger system sizes.   These results suggest that we have 
 QAHC phase in this range of $\theta$ at the filling number $\nu=3/4$.

  \end{widetext}

\end{document}